\documentclass[superscriptaddress,secnumarabic,nobibnotes,aps,
prd,showpacs,nofootinbib, twocolumn]{revtex4}
\usepackage{amsmath}
\usepackage{eurosym}
\usepackage{amssymb}
\usepackage{graphicx}
\usepackage{color}

\newcommand{\bea}{\begin{eqnarray}}
\newcommand{\eea}{\end{eqnarray}}
\newcommand{\be}{\begin{equation}}
\newcommand{\ee}{\end{equation}}

\newcommand{\nn}{\nonumber}

\begin{document}

\title{Spherically symmetric static vacuum solutions in hybrid
metric-Palatini gravity }
\author{Bogdan D$\check{\mathrm{a}}$nil$\check{\mathrm{a}}$}
\email{bogdan.danila22@gmail.com}
\affiliation{Astronomical Observatory, 19 Ciresilor Street, Cluj-Napoca, Romania,}
\affiliation{Department of Physics, Babes-Bolyai University, Kogalniceanu Street,
Cluj-Napoca 400084, Romania,}
\author{Tiberiu Harko}
\email{t.harko@ucl.ac.uk}
\affiliation{Department of Physics, Babes-Bolyai University, Kogalniceanu Street,
Cluj-Napoca 400084, Romania,}
\affiliation{School of Physics, Sun Yat-Sen University, Guangzhou 510275, People's
Republic of China}
\affiliation{Department of Mathematics, University College London, Gower Street, London
WC1E 6BT, United Kingdom,}
\author{Francisco S. N. Lobo}
\email{fslobo@fc.ul.pt}
\affiliation{Instituto de Astrof\'{\i}sica e Ci\^{e}ncias do Espa\c{c}o, Faculdade de
Ci\^encias da Universidade de Lisboa, Edif\'{\i}cio C8, Campo Grande,
P-1749-016 Lisbon, Portugal,}
\author{Man Kwong Mak}
\email{mankwongmak@gmail.com}
\affiliation{Departamento de F\'{\i}sica, Facultad de Ciencias Naturales, Universidad de
Atacama, Copayapu 485, Copiap\'o, Chile}
\date{\today }

\begin{abstract}
We consider vacuum static spherically symmetric solutions in the hybrid
metric-Palatini gravity theory, which is  a combination of the metric and
Palatini $f(R)$ formalisms unifying local constraints at the Solar
System level and the late-time cosmic acceleration. We adopt the
scalar-tensor representation of the hybrid metric-Palatini theory, in which
the scalar-tensor definition of the potential can be represented as a
Clairaut differential equation. Due to their mathematical complexity, it is
difficult to find exact solutions of the vacuum field equations, and therefore we adopt a numerical approach in studying the behavior of the metric functions and of the scalar field. After reformulating the field equations in a
dimensionless form, and by introducing a suitable independent radial
coordinate, the field equations are solved numerically. We detect the
formation of a black hole from the presence of  Killing
horizon for the time-like Killing vector in the metric
tensor components. Several models, corresponding to different functional
forms of the scalar field potential are considered. The thermodynamic properties of these black hole solutions (horizon temperature, specific heat, entropy and evaporation time due to Hawking luminosity) are also investigated in detail.
\end{abstract}

\maketitle
\tableofcontents

\section{Introduction}

The observational discovery of the recent acceleration of the Universe \cite%
{1n,2n,3n,4n,acc} has raised the fundamental theoretical problem if general
relativity, in its standard formulation, can fully account for all the
observed phenomena at both galactic and extra-galactic scales. The simplest
theoretical explanation for the observed cosmological dynamics consists in
slightly modifying the Einstein field equations, by adding to it a
cosmological constant $\Lambda $ \cite{Wein}. Together with the assumption of the existence of another mysterious component of the Universe, denoted dark matter \cite{dm1,dm2}, and assumed to be cold, the Einstein gravitational field equations can give an excellent fit to all observational data, thus leading to the formulation of the standard cosmological paradigm of our present days, called the $\Lambda $CDM model. However, despite its
apparent simplicity and naturalness, the introduction of the cosmological
constant raises a number of important questions for which no convincing
answers have been provided so far. Thus, the $\Lambda $CDM model can fit the observational data at a high level of precision, and despite being a very simple theoretical approach, no fundamental theory can explain it. Why is the cosmological constant so small and so fine-tuned? Why did the
Universe begin to accelerate so recently? And, after all, why is a cosmological constant necessary at all?

From a theoretical point of view two possible answers to the questions
raised by the recent acceleration of the Universe can be formulated. The
first, we may call the dark energy approach, assumes that the Universe is
filled by a mysterious and unknown component, called dark energy \cite%
{PeRa03, Pa03,8,Tsu}, which is fully responsible for the accelerated expansion of the Universe. The cosmological constant may correspond to a particular phase of the dynamical dark energy (ground state of a potential, let's say), and the
recent de Sitter phase may prove to be just an attractor of the dynamical
system describing the cosmological evolution. A second approach, the dark
gravity approach, assumes the alternative possibility that at large scales
the gravitational force may have a behavior different from that suggested by
standard general relativity. In the general relativistic description of
gravity, the starting point is the Hilbert-Einstein action, which can be
written down as $S=\int{\left(R/2\kappa ^2+L_m\right)\sqrt{-g} \; d^4x}$, where $R$ is the Ricci scalar, $\kappa $ is the gravitational coupling constant,
and $L_m$ is the matter Lagrangian, respectively. Hence, in dark gravity
theories for a full understanding of the gravitational interaction a
generalization of the Hilbert-Einstein action is necessary.

There are two possibilities to construct dark gravity theories. The first is
based on the modification of the geometric part of the Hilbert-Einstein
Lagrangian only. An example of such an approach is the $f(R)$ gravity
theory, introduced in \cite{Bu,Ba}, and in which the geometric part of the
action is generalized so that it becomes an arbitrary function $f(R)$ of the
Ricci scalar. Hence, in $f(R)$ gravity the total Hilbert-Einstein action can
be written as $S=\int{\left[f(R)/2\kappa ^2+L_m\right]\sqrt{-g} \; d^4x}$. The recent cosmological observations can be satisfactorily explained in the $f(R) $ theory, and a solution of the dark matter problem, interpreted as a
geometric effect in the framework of the theory, can also be obtained \cite%
{dm3}. In a more general approach one modifies both the geometric and the
matter terms in the Hilbert-Einstein action, thus allowing a coupling
between matter and geometry \cite{Bert,Harkof}. For reviews and in depth
discussions of $f(R)$ and other modified gravity theories see \cite%
{r1,r2,r3,r4,rn, r5,r6,r7,r8,r9}.

Einstein's general theory of relativity can be obtained by starting from two
different theoretical approaches, called the metric and the Palatini
formalisms \cite{Olmo}, respectively. When applied to the Hilbert-Einstein
action, these two approaches lead to the same gravitational field equations,
with the Palatini formalism also providing the explicit expression of the
symmetric connection in terms of the derivatives of the metric tensor.
However, in $f(R)$ gravity, as well as in other modified theories of
gravity, this does not happen anymore, and it turns out that the
gravitational field equations obtained using the metric approach are
generally different from those obtained by using the Palatini variation \cite{Olmo}. An important difference is related to the order of the field
equations, with the metric formulation usually leading to higher-order
derivative field equations, while in the Palatini approach the derived field
equations are always second order partial differential equations. On the
other hand, in the Palatini variational formulation a number of new
algebraic relations appear, which involve the matter fields and the
affine connection, such that the connection can be determined from a set of equations which couples it to the metric and to the matter fields.

Based on a hybrid combination of the metric and Palatini mathematical
formalisms, an extension of the $f(R)$ gravity theory was proposed in \cite{h1}. In this approach the (purely metric) Hilbert-Einstein action is
generalized by adding to it (metric-affine) correction terms obtained in the
spirit of the Palatini approach. Simple extensions of standard general
relativity with interesting properties can be constructed using both metric
and Palatini $f(R)$ theories. However, in each of these theories a
number of different pathological behaviors appear. Hence, by building a
bridge that relates these two apparently different approaches we may find a possibility of removing their individual failures.

A hybrid combination of the metric and Palatini formalisms was used in \cite{h1,h2} to construct a new type of gravitational Lagrangian. This gravitational theory is called hybrid metric-Palatini gravity (HMPG). From a
theoretical point of view the main result of this approach is that viable
gravity theories including elements of both formalisms can be obtained.
Moreover, in this class of theories it is possible to generate long-range
forces that do not contradict the classical local Solar System tests of
gravity. The analysis of the field equations and the construction of
solutions is greatly simplified with the use of the scalar-tensor
representation of the hybrid metric-Palatini theories. A simple example of
such a hybrid metric-Palatini theory can be constructed by adopting for the
gravitational Lagrangian the expression $R + f(\mathcal{R})$, where $%
\mathcal{R}$ is the Palatini scalar curvature. Such a gravitational action
maintains all the well-confirmed results of general relativity, which are
included in the Hilbert-Einstein part of the action $R$, and which describes
with a high precision gravitational phenomena at the scale of the Solar
System and of compact objects. On the other hand the metric-affine component $f(\mathcal{R})$ generates novel physical characteristics that may explain the recent cosmological observations of the accelerating Universe. In \cite{Ko1,Ko2} a similar formalism that interpolate between the metric and Palatini regimes was proposed for the study of $f(R)$ type theories. This approach is called the C-theory. In \cite{B1a} a generalization of HMPG was introduced.

The study of the cosmological and astrophysical implications of HMPG has attracted a lot of attention recently.
The properties of the Einstein static Universe in HMPG were studied in \cite{B}. Cosmological solutions obtained with the help of the scalar-tensor
representation of the theory were presented in \cite{c1}, were their
cosmological applications were also investigated. The cosmological field
equations were formulated as a dynamical system, and, by adopting some
specific functional forms of the effective scalar field potential, several
classes of cosmological solutions were obtained.
The  dynamical system approach was generalised in \cite{c5}, where new accelerating solutions that can be attractors in the phase space were found.
In \cite{c4} the evolution of the linear perturbations in HMPG
was considered. The full set of linearized evolution equations for the
perturbed physical and geometrical quantities were obtained, and it was
shown that important deviations from the $\Lambda$CDM model occur in the far
past, with ratio between the Newtonian potentials $\Phi $ and $\Psi$
presenting an oscillatory signature. Cosmological models were studied in
\cite{c6}. By using a combination of baryonic acoustic oscillations,
supernovae and cosmic microwave background data, the free parameters of the models can be constrained.
The analysis was further generalized using a specific HMPG model, given by $f (R )\propto R^2$ \cite{new1}, and the results were compared with the local constraints.

In the scalar-tensor representation of the HMPG theory,  new cosmological solutions were  obtained in \cite{new2} by either making an ansatz on the scale factor or on the effective potential. The efficiency of screening mechanisms in the hybrid metric-Palatini gravity was investigated in \cite{new3}. Bounds on the model were obtained using data from Solar System experiments, and  they can contribute to fix the range of viable hybrid gravity models. G\"{o}del-type solutions, in which  the matter source is a combination of a scalar with an electromagnetic fields, plus a perfect fluid were obtained in the framework of HMPG theory in \cite{new4}. The  existence of  G\"{o}del-type solutions indicates that HMPG does not solve the causal anomaly in the form of closed timelike curves that appears in general relativity.

HMPG also opens some new perspectives for the study of dark matter. The
virial theorem for galaxy clusters in HMPG was derived in \cite{c2}. It turns out that the total virial mass of the cluster is proportional to the effective
mass associated to the mass of the effective scalar field. Therefore, the
virial mass discrepancy in clusters of galaxies can be explained via the
geometric terms appearing in the generalized virial theorem. The HMPG dark matter model also predicts that the effects of the effective mass associated to the scalar field extend far beyond the virial radii of the clusters of galaxies. HMPG also allows for an explanation of the behavior of the rotational velocities of test particles gravitating around galaxies \cite{c3}. In the equivalent scalar-tensor description the rotational velocity can be obtained explicitly as a function of the scalar field. Hence all the
geometric and physical quantities, as well as the coupling constant in HMPG can be expressed as functions of measurable or observable parameters, such as, for example, the stellar dispersion velocity, the Doppler frequency shifts, the baryonic mass of the galaxy, and the tangential velocity, respectively.

The HMPG theory has also been explored in a plethora of other topics.
For instance, the problem of the well-posedness and the well-formulation of the Cauchy problem was discussed in \cite{P}. Wormhole solutions in HMPG have also been found in \cite{Lobo}, where it was
shown that these exotic geometries are supported by the higher order terms. Specific wormhole solutions in a generalized HMPG theory were  also found \cite{new5}. In these solutions the matter field obeys the null energy condition (NEC) everywhere, including the throat and up to infinity, so that there is no need for exotic matter.
In the context of compact objects, the internal structure and the physical properties of specific classes of neutron, quark and Bose-Einstein Condensate stars in HMPG were considered in \cite{B1}.
For reviews of HMPG theories, we refer the reader to \cite{rev} and \cite{book}, respectively.

Since Karl Schwarzschild obtained the first exact solution of the general
relativistic field equations in vacuum \cite{Sch1}, the search for black
hole type solutions describing the gravitational field outside massive
gravitating bodies proved to be of fundamental importance for the
theoretical understanding and observational testing of gravitational
theories (see \cite{0} for a review of the exact solutions of the Einstein
field equations). Black hole solutions allow testing the properties of the
gravitational force by using the electromagnetic emissivity properties of
thin disks that form around compact objects \cite{A1,
A2,A3,A4,A5,A6,A7,A8,A9,A10,A11,A12,A13,A14,A15}. For a review of the
possibilities of testing black hole candidates by using electromagnetic
radiation see \cite{RB}.

Many black hole type solutions have been obtained in different modified
theories of gravity, such as in brane world models \cite{Br0,Br1,Br2,Br3}, Eddington-inspired Born-Infeld \cite{Bh1}, higher derivative gravitational theory with a pair of complex conjugate ghosts \cite{Bh2}, de
Rham-Gabadadze-Tolley (dRGT) theory \cite{Bh3}, Gauss-Bonnet massive gravity coupled to Maxwell and Yang-Mills fields in five dimensions \cite{Bh4}, in the framework of the Poincar\'e gauge field theory with dynamical massless
torsion \cite{Bh5}, Rastall theory \cite{Bh8},
second-order generalized Proca theories with derivative vector-field
interactions coupled to gravity \cite{Bh10}, mimetic Born-Infeld gravity \cite{Bh11}, a class of vector-tensor theories of modified gravity \cite{Bh12}, and dilatonic dyon-like black hole solutions in a model with two Abelian gauge fields were also found \cite{Bh6}. Black hole solutions that can
accommodate both a nonsingular horizon and Yukawa asymptotics have been
considered in \cite{Bh9}.
In \cite{Bh7} it was shown that a large number of static, spherically symmetric metrics, which are regular at the origin, asymptotically flat, and have both an event
and a Cauchy horizon for a certain range of the parameters can be
interpreted as exact solutions of the Einstein equations coupled to ordinary
linear electromagnetism, that is, as sources of the Reissner-Nordstr\"{o}m
spacetimes.

In fact, the literature is extremely extensive, and we refer the reader to a solution-generating technique that maps a static charged
solution of the Einstein-Maxwell theory in four (or five) dimensions to a
five-dimensional solution of the Einstein-Maxwell-Dilaton theory \cite{Bh13}. Black hole solutions in Gauss-Bonnet-massive gravity in the presence
of power-Maxwell field were studied in \cite{Bh14,Bh14a,Bh14b,Bh14c}. In \cite{Bh15} it was
shown that black-hole solutions appear as a generic feature of the general
Einstein-scalar-Gauss-Bonnet theory with a coupling function $f(\phi)$. The
existing no-hair theorems are easily evaded for this model, and a large
number of regular black-hole solutions with scalar hair can be obtained.
The properties of black holes in static and spherically symmetric
backgrounds in U(1) gauge-invariant scalar-vector-tensor theories with
second-order equations of motion were studied in \cite{Bh16}. Exact asymptotically anti-de Sitter black hole solutions and
asymptotically Lifshitz black hole solutions with dynamical exponents $z = 0$
and $z = 4$ of four-dimensional conformal gravity coupled with a
self-interacting conformally invariant scalar field were obtained in \cite%
{Bh17}. The vacuum solutions around a spherically symmetric and static object in the Starobinsky model were studied with a perturbative approach in \cite{Bh18}. Dilatonic black holes in the presence of (non)linear electrodynamics have been studied in \cite{Bh19a,Bh19b}, respectively.

It is the main goal of the present paper to investigate the possibility of the existence of  spherically symmetric static vacuum solutions in the HMPG theory. In order to fulfil this goal,
we adopt the scalar-tensor representation of the theory, in which
the scalar-tensor definition of the potential can be represented as a
Clairaut differential equation. Even in the scalar-tensor representation, resembling the Brans-Dicke theory, the field equations show a high degree of mathematical complexity. Hence it turns out that it is extremely
difficult to find exact solutions of the vacuum gravitational field equations, and therefore for
the study of the behavior of the metric functions and of the scalar field one must
adopt numerical approaches.



The present paper is organized as follows. We briefly present the theoretical foundations and the field equations of HMPG in Section~\ref{sect2}. The field equations in  spherical symmetry for the vacuum case are written down in Section~\ref{sect3}, where their dimensionless formulation is introduced. Some general properties of the field equations are also discussed. The field equations are solved numerically in Section~\ref{sect4} for two particular choices of the scalar field potential, corresponding to a vanishing potential, and a Higgs-type potential, respectively. In each case, the behavior of the metric tensor coefficients and of the effective mass of the scalar field is considered in detail. The thermodynamic properties of the HMPG black holes are investigated in Section~\ref{sect5}, where the black hole temperature, specific heat, entropy, luminosity and life time are discussed. We discuss and conclude our results in Section~\ref{sect6}. In Appendix~\ref{Appa} we present the details of the transformations of the field equations to a dimensionless form.

\section{Field equations in HMPG}\label{sect2}

In the present Section we briefly review the action and the field equations
of HMPG. Its scalar-tensor formulation is presented, and the post-Newtonian parameters of the theory are also discussed.

\subsection{Action and gravitational field equations}

The action for HMPG can be constructed as \cite{h1}
\begin{equation}  \label{eq:S_hybrid}
S=\frac{1}{2\kappa^2}\int d^4 x \sqrt{-g} \left[ R + f(\mathcal{R})\right] +
S_m,
\end{equation}
where we have denoted $\kappa^2\equiv 8\pi G_0/c^4$, while $c$ and $G_0$ are
the standard speed of light and gravitational constant, respectively; $S_m$ is the matter action, defined as $S_m=\int d^4 x \sqrt{-g}\,\mathcal{L}_m $, where $\mathcal{L}_m$ is the matter Lagrangean; $R$ is
the metric Hilbert-Einstein term, and $\mathcal{R} \equiv g^{\mu\nu}\mathcal{%
R}_{\mu\nu} $ is the Palatini curvature. The tensor and $\mathcal{R}%
_{\mu\nu} $ is defined by using an independent connection $\hat{\Gamma}%
^\alpha_{\mu\nu} $ according to
\begin{equation}
\mathcal{R}_{\mu\nu} \equiv \hat{\Gamma}^\alpha_{\mu\nu ,\alpha} - \hat{%
\Gamma}^\alpha_{\mu\alpha , \nu} + \hat{\Gamma}^\alpha_{\alpha\lambda}\hat{%
\Gamma}^\lambda_{\mu\nu} -\hat{\Gamma}^\alpha_{\mu\lambda}\hat{\Gamma}%
^\lambda_{\alpha\nu}\,.
\end{equation}

In the following, the matter energy-momentum tensor $T_{\mu \nu}$ is defined as
\begin{equation}
T_{\mu\nu} \equiv -\left(\frac{2}{\sqrt{-g}}\right)\frac{ \delta (\sqrt{-g}%
\mathcal{L}_m)}{\delta g^{\mu\nu}}.
\end{equation}

After varying the action (\ref{eq:S_hybrid}) with respect to the metric, we
obtain the gravitational field equations of HMPG as
\begin{equation}  \label{efe}
G_{\mu\nu} + F(\mathcal{R})\mathcal{R}_{\mu\nu}-\frac{1}{2}f(\mathcal{R}%
)g_{\mu\nu} = \kappa^2 T_{\mu\nu},
\end{equation}
where we have denoted $F(\mathcal{R}) \equiv df(\mathcal{R})/d\mathcal{R}$.
After varying the action with respect to the independent connection one can easily show that the independent connection is compatible with the
metric $F(\mathcal{R})g_{\mu\nu}$, conformal to $g_{\mu\nu}$, with the
conformal factor given by $F(\mathcal{R})$. Hence we can obtain the field
equations in the equivalent form
\begin{eqnarray}  \label{ricci}
\mathcal{R}_{\mu\nu} &=& R_{\mu\nu} + \frac{3}{2}\frac{1}{F^2(\mathcal{R})}F(%
\mathcal{R})_{,\mu}F(\mathcal{R})_{,\nu}
\nonumber\\
&&-\frac{1}{F(\mathcal{R})}%
\nabla_\mu F(\mathcal{R})_{,\nu} - \frac{1}{2}\frac{1}{F(\mathcal{R})}%
g_{\mu\nu}\Box F(\mathcal{R}).
\end{eqnarray}
By taking the trace of the field equations (\ref{efe}) we obtain $\mathcal{R}
$ in terms of the trace $T$ of the matter energy-momentum tensor as
\begin{equation}\label{eq6}
F(\mathcal{R})\mathcal{R} -2f(\mathcal{R}) - R = \kappa^2 T.
\end{equation}

\subsection{Scalar-tensor formulation}

By introducing an auxiliary field $E$, the hybrid metric-Palatini action (\ref%
{eq:S_hybrid}) can be reformulated in the equivalent form of a scalar-tensor
theory, having the following action
\begin{equation}\label{eq7}
S=\frac{1}{2\kappa^{2}}\int\mathrm{d}^{4}x\sqrt{-g}[R+f(E)+f^{\prime }(E)(%
\mathcal{R}-E)],
\end{equation}
(for more technical details, we refer the
reader to \cite{h1}).

As one can easily see, for $E=\mathcal{R}$, the action given by Eq.~(\ref{eq7}) reduces to the action (\ref{eq:S_hybrid}). Hence, it turns out that if $f^{\prime\prime}(\mathcal{R})\neq0$, the field $E$ is
dynamically equivalent to the Palatini scalar $\mathcal{R}$. By introducing
the definitions $\;$%
\begin{equation}
\phi\equiv f^{\prime}(E), \qquad V(\phi)=Ef^{\prime}(E)-f(E),  \label{lagr}
\end{equation}
the action takes the form
\begin{equation}  \label{eq:S_scalar1}
S= \frac{1}{2\kappa^2}\int d^4 x \sqrt{-g} \left[ R + \phi\mathcal{R}-V(\phi)%
\right] +S_m  .
\end{equation}
If we vary this action with respect to the metric, the scalar $\phi$ and the
connection, respectively, we obtain the following field equations
\begin{eqnarray}
R_{\mu\nu}+\phi \mathcal{R}_{\mu\nu}-\frac{1}{2}\left(R+\phi\mathcal{R}%
-V\right)g_{\mu\nu}&=&\kappa^2 T_{\mu\nu} ,  \label{eq:var-gab} \\
\mathcal{R}-V_\phi&=&0 \,,  \label{eq:var-phi} \\
\hat{\nabla}_\alpha\left(\sqrt{-g}\phi g^{\mu\nu}\right)&=&0 , \
\label{eq:connection}
\end{eqnarray}
respectively.

It is interesting to mention that Eq. (\ref{lagr}) is a Clairaut
differential equation \cite{Claraut}, that is, it has the form
\begin{equation}
Ef^{\prime }(E)-f(E)=V\left( f^{\prime }\left( E\right) \right) \,.
\label{fe.04}
\end{equation}%
This equation has a linear general solution given by
\begin{equation}
f( E) = h\, E-V( h) \,,  \label{fe.05}
\end{equation}
where $h$ is a constant, or, equivalently,
\be
f\left( \mathcal{R}\right) = h\, \mathcal{R}-V( h),
\ee
and a singular solution, which can be found from the differential equation
\begin{equation}
\frac{\partial V\left( f^{\prime }\left( E\right) \right) }{\partial
f^{\prime }}-E=0 \,.  \label{fe.06}
\end{equation}

Hence in our mathematical formalism for the nonsingular solution of the Clairaut equtation the function $f\left(\mathcal{R}\right)$ is {\it a linear function of the Palatini curvature}. In this case for the vacuum state with $T=0$ the trace equation (\ref{eq6}) becomes
\be
-h\mathcal{R}-R+2V(h)=0.
\ee
With the use of the non-singular solution (\ref{fe.05}) we can express the potential $V(\phi)$ of the effective scalar field as
\be
V(\phi)=\left(\phi-h\right)E+V(h),
\ee
giving
\be\label{eq18}
E=\mathcal{R}=\frac{V(\phi)-V(h)}{\phi-h}.
\ee

It is interesting to note that when $V(h)>>\left(\phi-h\right)E$, the scalar field generates an effective cosmological constant, whose numerical value is  determined by the functional form of the potential estimated for a constant value of the scalar field.

For a zero scalar field potential, $V\left(\phi \right)\equiv 0$, from Eq.~(\ref{eq18}) it follows that $\mathcal{R}=0$, also giving $R=0$. 

For a potential of the form $V(\phi)=-\left(\mu ^2/2\right)\phi^2+\left(\zeta /4\right)\phi ^4$, we obtain
\be
\mathcal{R}=\frac{-\mu ^2\left(\phi ^2-h^2\right)/2+\zeta \left(\phi ^4-h^4\right)/4}{\phi-h},
\ee
\bea
R&=&-h\frac{-\mu ^2\left(\phi ^2-h^2\right)/2+\zeta \left(\phi ^4-h^4\right)/4}{\phi-h}\nonumber\\
&&-\mu^2 h^2+\frac{\zeta}{2}h^4,
\eea
\bea
f\left(\mathcal{R}\right)&=&h\frac{-\mu ^2\left(\phi ^2-h^2\right)/2+\zeta \left(\phi ^4-h^4\right)/4}{\phi-h}\nonumber\\
&&+\frac{\mu^2}{2} h^2-\frac{\zeta}{4}h^4.
\eea

Once the variation of $\phi$ is known from the solution of the gravitational field equations, the Palatini scalar curvature, the metric Hilbert scalar curvature as well as the function $f\left(\mathcal{R}\right)$ can be reconstructed directly, with the trace equation (\ref{eq6}) giving the metric Hilbert curvature, while the functional form of $f\left(\mathcal{R}\right)$ follows directly from Eq.~(\ref{fe.05}).

On the other hand, the solution of Eq.~(\ref{eq:connection}) shows that the
independent connection is the Levi-Civita connection of the metric $%
h_{\mu\nu}=\phi g_{\mu\nu}$. Thus, HPMG is a bi-metric theory, with $%
\mathcal{R}_{\mu\nu}$ and $R_{\mu\nu}$ related by
\begin{equation}  \label{eq:conformal_Rmn}
\mathcal{R}_{\mu\nu}=R_{\mu\nu}+\frac{3}{2\phi^2}\partial_\mu \phi
\partial_\nu \phi-\frac{1}{\phi}\left(\nabla_\mu \nabla_\nu \phi+\frac{1}{2}%
g_{\mu\nu}\Box\phi\right)  .
\end{equation}
Therefore the two Ricci scalars are related as
\begin{equation}
\mathcal{R}=R+\frac{3}{2\phi^2}\partial_\mu \phi \partial^\mu \phi-\frac{3}{%
\phi}\Box \phi.
\end{equation}
With the help of this relation  we eliminate now in action (\ref{eq:S_scalar1}) the independent connection. Thus we obtain the following scalar-tensor representation of HMPG \cite{h1},
\be
S=S_g\left(g,\phi\right)+S_m,
\ee
where $S_g\left(g,\phi\right)$ is given by
\begin{eqnarray}  \label{eq:S_scalar2}
S_g\left(g,\phi\right)&=& \frac{1}{2\kappa^2}\int d^4 x \sqrt{-g} \Big[ (1+\phi)R
	\nonumber  \\
&&+\frac{3}{2\phi}%
\partial_\mu \phi \partial^\mu \phi -V(\phi)\Big].
\end{eqnarray}

Despite some superficial analogies, this action essentially differs in their couplings of the scalar to the curvature from the $w=-3/2$
Brans-Dicke theory action. However, it belongs to the class of general Bergmann-Wagoner-Nordtvedt scalar-tensory theories \cite{N1, N2, N3}, whose action  for the vacuum state is given by
\be\label{20}
S=\frac{1}{2\kappa ^2}\int{\left[f\left(\phi \right)R+g\left(\phi\right)\partial_{\mu}\phi \partial ^{\mu}\phi+\lambda \left(\phi\right)\right]\sqrt{-g}d^4x},
\ee
where $f(\phi)$, $g(\phi)$ and $\lambda (\phi)$ are arbitrary functions of the scalar field $\phi$. A comparison with the action (\ref{eq:S_scalar2}) of the HMPG theory in the scalar-tensor representation shows that its action is indedd of Bergmann-Wagoner-Nordtvedt type, with $f(\phi)=1+\phi$,  $g(\phi)=3/2\phi$, and $\lambda (\phi)=-V(\phi)$, respectively. An important property of the Bergmann-Wagoner-Nordtvedt theories is that by means of the transformations \cite{N2}
\be\label{21}
\bar{g}_{\mu \nu}=f(\phi)g_{\mu\nu}, \frac{d\phi}{d\bar{\phi}}=f(\phi)\left[f(\phi)g(\phi)-\frac{3}{2}\left(\frac{df(\phi)}{d\phi}\right)^2\right],
\ee
the action (\ref{20}) can be transformed to the form
\be
S=\frac{1}{2\kappa ^2}\int{\left[\bar{R}-n\bar{g}^{\mu \nu}\partial _{\mu}\bar{\phi}\partial_{\nu}\bar{\phi}+2\lambda\left(\bar{\phi}\right)\right]\sqrt{-\bar{g}}d^4x},
\ee
where $n=\pm 1$. {\it A crucial mathematical requirement for transformations} (\ref{21}) to be valid is that they must be {\it nonsingular} for the considered range of the scalar field variable.

Let's apply now the transformations (\ref{21}) to the action (\ref{eq:S_scalar2}) of the HMPG theory. We introduce first the conformal transformation of the metric $\bar{g}_{\mu\nu}=\left(1+\phi\right)g_{\mu\nu}$, which transforms the action (\ref{eq:S_scalar2}) to the Einstein frame form
\be
S=\frac{1}{2\kappa ^2}\int{\left[\bar{R}+\frac{3}{2\phi}\frac{\bar{g}^{\mu \nu}\partial _{\mu}\phi \partial_{\nu}\phi}{\left(1+\phi\right)^2}-\frac{V(\phi)}{\left(1+\phi\right)^2}\right]\sqrt{-\bar{g}}d^4x}.
\ee

Next, we introduce the scalar field transformation
\be\label{22}
\phi=\tan^2 \left(\sqrt{\frac{3}{8}}\;\bar{\phi}\right),
\ee
which follows from the second equation in (\ref{21}). This transformation will transform the HMPG vacuum theory into a canonical scalar field theory with a very specific potential. However, the transformation $\phi\rightarrow \bar{\phi}$ given by Eq.~(\ref{22}) is {\it singular}, with $\phi =0$ for $\bar{\phi}=\sqrt{8/3}k\pi$, $k=0,1,2,3...$, and $\phi \rightarrow \infty$ for $\bar{\phi}\rightarrow \sqrt{8/3}\left[\left(-1\right)^k\pi/2+2k\pi\right]$, $k=0,1,2,...$. Hence, even that one could find the solution of the vacuum field equations of the HMPG theory in the canonical scalar field representation in the Einstein frame, there is no guarantee that the obtained solution would generate mathematically consistent and well behaved solutions of the field equations in the Jordan frame, in which the HMPG theory is naturally formulated.

For the sake of comparison we will also briefly present the case of the standard Brans-Dicke theory \cite{Dicke1, Dicke2}, with the vacuum gravitational action given by
\be
S=\frac{1}{2\kappa ^2}\int{\left(\phi R-\frac{\omega}{\phi}g^{\mu \nu}\partial _{\mu}\phi \partial _{\nu}\phi\right)\sqrt{-g}d^4x},
\ee
which with the help of the transformations
\be\label{26}
\bar{g}_{\mu \nu}=\phi g_{\mu \nu}, \phi =e^{\bar{\phi}},
\ee
can be transformed into the canonical form
\be
S=\frac{1}{2\kappa^2}\int{\left[\bar{R}-\left(\omega +\frac{3}{2}\right)\bar{g}^{\mu \nu}\partial _{\mu}\bar{\phi}\partial _{\nu}\bar{\phi}\right]\sqrt{-\bar{g}}d^4x}.
\ee

As one can easily see from Eq.~(\ref{26}), the transformation law $\phi\rightarrow \bar{\phi}$ of the scalar field in the Brans-Dicke theory is {\it nonsingular}, except for $\bar{\phi}\rightarrow \pm \infty$. This makes the scalar field transformation mathematical properties in standard Brans-Dicke theory different as compared to the transformation (\ref{22}) of the scalar field in HMPG theory.

By substituting Eq.~(\ref{eq:var-phi}) and Eq.~(\ref{eq:conformal_Rmn}) in
Eq.~(\ref{eq:var-gab}), we can write the metric field equation as an
effective Einstein field equation given by
\begin{equation}
G_{\mu\nu}=\kappa^2 T^{\mathrm{eff}}_{\mu\nu},
\end{equation}
where the effective energy-momentum tensor is defined according to
\begin{eqnarray}
T^{\mathrm{eff}}_{\mu\nu}&=&\frac{1}{1+\phi} \Bigg\{ T_{\mu\nu} - \frac{1}{%
\kappa^2} \Bigg[ \frac{1}{2}g_{\mu\nu}\left(V+2\Box\phi\right)+
\nabla_\mu\nabla_\nu\phi
\nonumber\\
&&-\frac{3}{2\phi}\partial_\mu \phi \;\partial_\nu
\phi + \frac{3}{4\phi}g_{\mu\nu}(\partial \phi)^2 \Bigg] \ \Bigg\} .
\label{effSET}
\end{eqnarray}

The scalar field is governed by an effective Klein-Gordon type second-order
evolution equation, given by
\begin{equation}  \label{eq:evol-phi}
-\Box\phi+\frac{1}{2\phi}\partial_\mu \phi \partial^\mu \phi+\frac{\phi[%
2V-(1+\phi)V_\phi]} {3}=\frac{\phi\kappa^2}{3}T\,,
\end{equation}
(we refer the reader to \cite{h1} for more
details on the derivation of this equation).
The Klein-Gordon evolution equation indicates that, unlike in the
Brans-Dicke ($w=-3/2$) case, in the present theory the scalar field is
dynamical. Therefore, the theory does not experience the microscopic
instabilities that emerge in Palatini models with infrared corrections \cite%
{Olmo}. As for the matter energy-momentum tensor, it turns out that it is
independently conserved, and hence it satisfies the standard condition $%
\nabla _{\mu }T^{\mu}_{\nu}=0$. 

\subsection{The post-Newtonian parameters}

The post-Newtonian parameters of gravitational theories are important
indicators that help us to determine the viability of the theory by using
local gravitational tests. In this respect we consider the post-Newtonian
analysis of HMPG, where we perturb Eqs. (\ref{effSET}) and (\ref{eq:evol-phi})
in a Minkowskian background. We introduce first a quasi-Minkowskian
coordinate system, in which $g_{\mu\nu}\approx \eta_{\mu\nu}+h_{\mu\nu}$,
with $|h_{\mu\nu}|\ll 1$, and we take $\phi=\phi_0+\varphi(x)$, where $%
\phi_0 $ denotes the asymptotic value of the field far away from the
gravitating sources. Hence for this class of theories we can obtain the
standard post-Newtonian metric up to second order, together with the
following expressions of the relevant astrophysical parameters (we refer the
reader to \cite{rev} for details)
\begin{eqnarray}
G_{\mathrm{eff}}&\equiv & \frac{\kappa^2}{8\pi (1+\phi_0)}\left(1+\frac{%
\phi_0}{3}e^{-m_\varphi r}\right), \\
\gamma &\equiv & \frac{\left[1+\phi_0\exp \left(-m_{\varphi} r\right)/3%
\right]}{\left[1-\phi_ 0\exp \left(-m_{\varphi } r\right)/3\right]},
\label{gamma0} \\
m_\varphi^2 &\equiv &\frac{1}{3} \left[ 2V-V_{\phi}-\phi(1+\phi)V_{\phi\phi}%
\right]\big|_{\phi=\phi_0}\,.  \label{mass}
\end{eqnarray}
In HMPG there are two possibilities to obtain the value $\gamma\approx 1$ of
the PPN parameter $\gamma $. The first one is identical to the one used
in the $f(R)$-type theories, and requires the existence of a very massive
scalar field \cite{rn}. The second possibility consists in imposing a very
small background scalar field $\phi_0\ll 1$, so that regardless of the
magnitude of $m_\varphi$, the Yukawa-type corrections are very small. This
latter case leaves the local gravity tests unaffected, but it allows for the
existence of a long-range scalar field that can modify the cosmological
dynamics of the Universe.

\section{Spherically symmetric vacuum field equations in HMPG}\label{sect3}

\subsection{Metric and field equations}

In the following, we assume that the geometry outside gravitating objects can
be represented by the following line element in curvature coordinates,
\begin{equation}
ds^2=-e^{\nu(r)}c^2dt^2 + e^{\lambda(r)} dr^2 + r^2 \left(d \theta^2 + \sin
^2\theta d\varphi^2 \right),  \label{whmetric}
\end{equation}
representing a static and spherically symmetric space-time. The metric
functions $\nu(r)$ and $\lambda(r)$ are functions of the radial coordinate $%
r $ only, with the range $0 \leq r < \infty$. At least theoretically, in the
framework of HMPG we can construct asymptotically flat spacetimes, in which $%
\nu(r) \rightarrow 0$ and $\lambda(r) \rightarrow 0$ as $r \rightarrow
\infty $. In the following, our main goal is to investigate vacuum solutions
of the HMPG theory, for which all the components
of the energy-momentum tensor vanish identically, $T_{\mu \nu}\equiv 0$.

Using the metric (\ref{whmetric}), the effective Einstein field equation (%
\ref{effSET}) provides the following set of vacuum gravitational field
equations
\begin{eqnarray}
\frac{1}{r^{2}}\left[ 1-e^{-\lambda }\left( 1-r\lambda ^{\prime }\right) %
\right] (1+\phi )-e^{-\lambda }\left( \phi ^{\prime \prime }-\frac{3\phi
^{\prime 2}}{4\phi }\right)
	\nonumber \\
+\frac{\phi ^{\prime }}{2r}e^{-\lambda }\left(
r\lambda ^{\prime }-4\right) -\frac{V(\phi )}{2}=0,  \label{rho}
\end{eqnarray}
\begin{eqnarray}
&&\left[ \frac{1}{r^{2}}(e^{-\lambda }-1)+\frac{\nu ^{\prime }}{r}e^{-\lambda }%
\right] (1+\phi )
	\nonumber \\
&&+\phi ^{\prime }\left( \frac{\nu ^{\prime }}{2}+\frac{2}{r}+%
\frac{3\phi ^{\prime }}{4\phi }\right) e^{-\lambda }+\frac{V(\phi )}{2}=0\,,
\label{hybrid_pr}
\end{eqnarray}%
\begin{eqnarray}
\Bigg[\left( \frac{\nu ^{\prime \prime }}{2}+\left( \frac{\nu ^{\prime }}{2%
}\right) ^{2}+\frac{\nu ^{\prime }}{2r}\right) e^{-\lambda }-\frac{1}{2}%
\frac{\lambda ^{\prime }e^{-\lambda }}{r}\left( 1+r\frac{\nu ^{\prime }}{2}%
\right) \Bigg]
	\nonumber  \\
\times (1+\phi ) +\left[ \phi ^{\prime \prime }+\frac{\phi ^{\prime
}\nu ^{\prime }}{2}+\frac{3\phi ^{\prime 2}}{4\phi }\right] e^{-\lambda }
	\notag \\
+\frac{\phi ^{\prime }}{r}e^{-\lambda }\left( 1-\frac{r\lambda ^{\prime }}{2%
}\right) +\frac{V(\phi )}{2}=0,
 \label{pt}
\end{eqnarray}
where a prime denotes the derivative with respect to the radial coordinate $%
r $. The effective vacuum Klein-Gordon equation (\ref{eq:evol-phi}) is given
by
\bea\label{modKGeq0}
-\left[ \phi ^{\prime \prime }+\frac{\phi ^{\prime }\nu ^{\prime }}{2}-\frac{%
\phi ^{\prime 2}}{2\phi }+\frac{2\phi ^{\prime }}{r}\right] e^{-\lambda }+%
\frac{\phi ^{\prime }\lambda ^{\prime }}{2}e^{-\lambda }
	\nonumber\\
+\frac{\phi }{3}\left[ 2V(\phi )-(1+\phi )V_{\phi }(\phi )\right] =0.
\eea

Note that once the functional dependence of the scalar field potential $%
V(\phi (r))$ is given, Eqs.~(\ref{rho})-(\ref{modKGeq0}) provide four
independent ordinary nonlinear differential equations for three unknown quantities, $\nu (r)$, $\lambda (r)$%
, and $\phi (r)$, respectively. However, similarly to the case of standard general relativity, Eq.~(\ref{pt}) is a consequence of the two other field equations, and of the generalized Klein-Gordon equation. Therefore, in order to investigate the black hole properties in HPMG it is enough to consider the solutions of the system formed of Eqs.~(\ref{rho}), (\ref{hybrid_pr}) and (\ref{modKGeq0}).

\subsection{The mass function and the dynamical system formulation}

In order to simplify the mathematical formalism we introduce a new function $%
m_{eff}(r)$, and we redefine the metric tensor component $e^{-\lambda }$ as
\begin{equation}
e^{-\lambda }=1-\frac{2Gm_{eff}(r)}{c^{2}r},
\end{equation}%
so that
\begin{equation}
\mathbf{\ }\lambda ^{\prime }e^{-\lambda }=\frac{2G}{c^{2}r}\left(
m_{eff}^{\prime }-\frac{m_{eff}}{r}\right) .
\end{equation}

Then the basic equations describing the vacuum metric tensor components in HMPG can be written as
\begin{equation}
\frac{d\phi }{dr}=u,  \label{eq1}
\end{equation}
\begin{eqnarray}  \label{eq2}
\frac{dm_{eff}}{dr}=\frac{c^{2}r^{2}}{2G(1+\phi +ur/2)}\Big[ \left(
1-2Gm_{eff}/c^{2}r\right) \times
	\nn \\
\times \left( u^{\prime }-3u^{2}/4\phi +2u/r\right) -
\frac{2Gm_{eff}}{c^{2}r^{3}}\left( 1+\phi \right) +V/2\Big]
	\nn \\
+\frac{m_{eff}}{r},
\end{eqnarray}
\begin{equation}  \label{eq3}
\nu ^{\prime }=\frac{\frac{1}{r}-\left[ \frac{u\left( 8\phi +3ur\right) }{%
4\phi \left( 1+\phi \right) }+\frac{1}{r}\right] \left( 1-\frac{2Gm_{eff}}{%
c^{2}r}\right) -\frac{rV\left( \phi \right) }{2\left( 1+\phi \right) }}{%
\left( 1-\frac{2Gm_{eff}}{c^{2}r}\right) \left[ 1+\frac{ur}{2\left( 1+\phi
\right) }\right] },
\end{equation}
\begin{eqnarray}
\nu ^{\prime \prime }&=&\frac{2G}{c^{2}r}\frac{\left( m_{eff}^{\prime }-%
\frac{m_{eff}}{r}\right) \left( 1+r\frac{\nu ^{\prime }}{2}\right) }{\left(
1-\frac{2Gm_{eff}}{c^{2}r}\right) }-\frac{u\left(5ur-8\phi\right)}{2r\phi
(1+\phi )}
	\nn \\
&&-\frac{2u}{r(1+\phi )}
-\frac{1}{(1+\phi )\left( 1-\frac{2Gm_{eff}}{c^{2}r}\right) }\times
	\notag \\
&& \left\{ \frac{%
2\phi }{3}\left[ 2V-(1+\phi )V_{\phi }\right] +V\right\}  -  \frac{\nu ^{\prime 2}}{2}-\frac{\nu ^{\prime }}{r} ,  \label{eq4}
\end{eqnarray}
\begin{eqnarray}
u^{\prime }=\frac{\frac{Gu}{c^{2}r}\left( m_{eff}^{\prime }-\frac{m_{eff}}{r}%
\right) +\frac{\phi }{3}\left[ 2V-(1+\phi )V_{\phi }\right] }{1-\frac{%
2Gm_{eff}}{c^{2}r}}
	\nn \\
-\frac{u\nu ^{\prime }}{2}+\frac{u^{2}}{2\phi }-\frac{2u}{%
r}.  \label{eq5}
\end{eqnarray}

To obtain Eq. (\ref{eq4}) we have proceeded as follows: we first rewrite the generalized Klein-Gordon equation (\ref{modKGeq0}) as
\bea
\left( \phi ^{\prime \prime }+\frac{\phi ^{\prime }\nu ^{\prime }}{2}\right)
e^{-\lambda }-\frac{\phi ^{\prime }\lambda ^{\prime }}{2}e^{-\lambda
}=\left( \frac{\phi ^{\prime 2}}{2\phi }-\frac{2\phi ^{\prime }}{r}\right)
e^{-\lambda }
	\nonumber\\
+\frac{\phi }{3}\left[ 2V(\phi )-(1+\phi )V_{\phi }(\phi )%
\right] .
\eea
Substituting the left hand side of this equation into Eq. (\ref{pt}), we find
\begin{eqnarray}
\Bigg[\left( \frac{\nu ^{\prime \prime }}{2}+\left( \frac{\nu ^{\prime }}{2%
}\right) ^{2}+\frac{\nu ^{\prime }}{2r}\right) e^{-\lambda }-\frac{1}{2}%
\frac{\lambda ^{\prime }e^{-\lambda }}{r}\left( 1+r\frac{\nu ^{\prime }}{2}%
\right) \Bigg]
	\nonumber\\
+\frac{5\phi ^{\prime 2}}{4\phi (1+\phi )}e^{-\lambda }+
- \frac{\phi ^{\prime }}{r(1+\phi )}e^{-\lambda }+\frac{\phi }{3(1+\phi )}\times
	\nonumber\\
\times \left[ 2V(\phi )-(1+\phi )V_{\phi }(\phi )\right] +\frac{V(\phi )}{2(1+\phi )%
}=0.
\end{eqnarray}
We multiply now with $e^{\lambda }$ to obtain
\begin{eqnarray}\label{pt1}
\Bigg[\left( \frac{\nu ^{\prime \prime }}{2}+\left( \frac{\nu ^{\prime }}{2%
}\right) ^{2}+\frac{\nu ^{\prime }}{2r}\right) -\frac{1}{2}\frac{\lambda
^{\prime }}{r}\left( 1+r\frac{\nu ^{\prime }}{2}\right) \Bigg]
	\nonumber\\
+\frac{5\phi^{\prime 2}}{4\phi (1+\phi )}-
\frac{\phi ^{\prime }}{r(1+\phi )}+\frac{1}{(1+\phi )}\times
	\nonumber\\
\times\Bigg\{ \frac{\phi }{%
3}\left[ 2V(\phi )-(1+\phi )V_{\phi }(\phi )\right] +
\frac{V(\phi )}{2}%
\Bigg\} e^{\lambda }=0.
\end{eqnarray}

Expressing $\nu ^{\prime \prime }$ from the above equation leads directly to
Eq. (\ref{eq4}).

\subsection{ The dimensionless form of the field equations}

In order to simplify the mathematical and numerical formalism, we introduce now a set of dimensionless variables $\left( \eta,M_{eff},U,v\right) $, defined as
\bea\label{dimvar}
\hspace{-0.5cm}r&=&\frac{2GM_{\odot}}{c^{2}}n\eta , \qquad  m_{eff}=nM_{\odot}M_{eff}\left( \eta
\right) ,
	\nonumber\\
\hspace{-0.5cm}u&=&\frac{c^{2}}{2GM_{\odot}n}U\left( \eta \right) ,
	V\left( \phi \right) = 2\left( \frac{c^{2}}{2GM_{\odot}n}\right) ^{2}v\left( \phi \right) .
\eea

The explicit representation of the physical and geometrical quantities in the new variables is represented in Appendix~\ref{Appa}.

Hence the system of Eqs. (\ref{eq1})-(\ref{eq5}) takes the dimensionless
form

\begin{equation}
\frac{d\phi }{d\eta }=U,  \label{46}
\end{equation}

\begin{eqnarray}
\frac{dM_{eff}}{d\eta }=\frac{\eta ^{2}}{1+\phi +\eta U/2}
\Big\{ \left[ \left( 1-M_{eff}/\eta
\right) \right] \big [ dU/d\eta
	\nn \\
-3U^{2}/4\phi +2U/\eta \big] -\frac{M}{%
\eta ^{3}}\left( 1+\phi \right) +v\Big\}
+\frac{M_{eff}%
}{\eta },  \label{47}
\end{eqnarray}
\begin{equation}
\frac{d\nu }{d\eta }=\frac{\frac{1}{\eta }-\left\{ \frac{U\left( \eta
\right) \left[ 8\phi +3\eta U\left( \eta \right) \right] }{4\phi \left(
1+\phi \right) }+\frac{1}{\eta }\right\} \left[ 1-\frac{M_{eff}\left( \eta
\right) }{\eta }\right] -\frac{\eta v\left( \phi \right) }{1+\phi }}{\left(
1-\frac{M_{eff}\left( \eta \right) }{\eta }\right) \left[ 1+\frac{\eta
U\left( \eta \right) }{2\left( 1+\phi \right) }\right] },  \label{48}
\end{equation}%
\begin{eqnarray}\label{49}
&&\frac{d^{2}\nu }{d\eta ^{2}} =\frac{1}{\eta }\frac{\left( 1+%
\frac{\eta }{2}\frac{d\nu }{d\eta }\right) \left( \frac{dM_{eff}}{d\eta }-%
\frac{M_{eff}}{\eta }\right) }{\left( 1-\frac{M_{eff}}{\eta }\right) }
	\nn \\
&&-\frac{5U(\eta)^2}{2 \phi (1+\phi )}+\frac{2 U(\eta)}{\eta (1+\phi
)}   -\frac{1}{2}\left(
\frac{d\nu }{d\eta }\right) ^{2}-\frac{1}{\eta }\frac{d\nu }{d\eta }
	\nonumber\\
&&-\frac{1}{(1+\phi )\left( 1-\frac{M_{eff}}{\eta }\right) }\left\{ \frac{%
2\phi }{3}\left[ 2V-(1+\phi )V_{\phi }\right] +V\right\} ,
\nn \\
\end{eqnarray}

\begin{eqnarray}
&&\frac{dU\left( \eta \right) }{d\eta }=
-\frac{U\left( \eta \right) }{2}\frac{d\nu }{d\eta }+\frac{%
U^{2}\left( \eta \right) }{2\phi }-\frac{2U\left( \eta \right) }{\eta }+
	\nn \\
&&\frac{\frac{U\left( \eta \right) }{%
2\eta }\left[ \frac{dM_{eff}\left( \eta \right) }{d\eta }-\frac{%
M_{eff}\left( \eta \right) }{\eta }\right] +\frac{2\phi }{3}\left[ 2v(\phi
)-(1+\phi )v_{\phi }(\phi )\right] }{1-\frac{M_{eff}\left( \eta \right) }{%
\eta }}.
	\nn \\
\label{50}
\end{eqnarray}

We introduce now a new variable $\xi =1/\eta $, so that
\begin{equation}
\frac{1}{r}=\frac{c^{2}}{2GM_{\odot}n}\frac{1}{\eta }=\frac{c^{2}}{%
2GM_{\odot}n}\xi .
\end{equation}

When $r\rightarrow \infty $, $\xi \rightarrow 0$, while for $r\rightarrow 0$, we have $\xi \rightarrow \infty $. In the new variable, Eqs. (\ref{46})-(\ref{50}) take the form
\begin{widetext}
\begin{equation}
\frac{d\phi }{d\xi }=-\frac{1}{\xi ^{2}}U,  \label{53}
\end{equation}%
\begin{equation}
\frac{dM_{eff}}{d\xi }=\frac{\left( 1-M_{eff}\xi \right) \left[ \xi
^{2}dU/d\xi +3U^{2}/4\phi -2\xi U\right] +M_{eff}\xi ^{3}\left( 1+\phi
\right) -v}{\xi ^{4}\left( 1+\phi +U/2\xi \right) }-\frac{M_{eff}}{\xi }.
\label{54}
\end{equation}
\begin{equation}
\frac{d\nu }{d\xi }=-\frac{\xi -\left\{ \frac{U\left( \xi \right) \left[
8\phi +3U\left( \xi \right) /\xi \right] }{4\phi \left( 1+\phi \right) }+\xi
\right\} \left[ 1-\xi M_{eff}\left( \xi \right) \right] -\frac{v\left( \phi
\right) }{\xi \left( 1+\phi \right) }}{\xi ^{2}\left[ 1-\xi M_{eff}\left(
\xi \right) \right] \left[ 1+\frac{U\left( \xi \right) }{2\xi \left( 1+\phi
\right) }\right] },  \label{55}
\end{equation}%
\begin{eqnarray}\label{56}
\frac{d^{2}\nu }{d\xi ^{2}} &=&\frac{\left( 1-%
\frac{\xi }{2}\frac{d\nu }{d\xi }\right) \left(\xi \frac{ dM_{eff}}{d\xi }-%
M_{eff}\right) }{\left( 1-\xi M_{eff}\right) }-\frac{%
5 U(\xi)^2 }{2 \xi^4 \phi (1+\phi )}+\frac{2u}{\xi^3 (1+\phi
)}-\nonumber\\
&&\frac{1}{\xi^4 (1+\phi )\left( 1-\xi M_{eff}\right) }\left\{ \frac{%
2\phi }{3}\left[ 2V-(1+\phi )V_{\phi }\right] +V\right\} -\frac{1}{2}\left(
\frac{d\nu }{d\eta }\right) ^{2}-\frac{1}{\xi}\frac{d\nu }{d\xi },
\end{eqnarray}
\begin{equation}\label{57}
\frac{dU\left( \xi \right) }{d\xi }=\frac{\frac{\xi ^{2}U\left( \xi \right)
}{2}\left[ \xi \frac{dM_{eff}\left( \xi \right) }{d\xi }+M_{eff}\left( \xi
\right) \right] -\frac{2\phi }{3}\left[ 2v(\phi )-(1+\phi )v_{\phi }(\phi )%
\right] }{\xi ^{2}\left[ 1-\xi M_{eff}\left( \xi \right) \right] }-\frac{%
U\left( \xi \right) }{2}\frac{d\nu }{d\xi }-\frac{1}{\xi ^{2}}\frac{%
U^{2}\left( \xi \right) }{2\phi }+\frac{2U\left( \xi \right) }{\xi }.
\end{equation}
\end{widetext}

In their dimensionless form in $\xi $ the field equations must be solved
with the fixed initial conditions
\begin{equation}  \label{58}
M_{eff}\left( 0\right) =1, \qquad  \nu (0)=0, \qquad  \nu ^{\prime }(0)=0,
\end{equation}%
and arbitrary numerical values for $u(0)=u_0$ and $\phi \left( 0\right) =\phi _0$.

\subsection{General properties of the gravitational field equations}

In order to simplify our formalism, we represent the metric tensor coefficient $e^{\nu }$ as
\begin{equation}
e^{\nu (r)}=\Psi \left( \phi (r)\right) e^{\beta (r)},
\end{equation}%
where $\Psi \left( \phi (r)\right) $ and $\beta (r)$ are functions to be
determined from the gravitational field equations. Then we immediately find
\begin{equation}
\nu ^{\prime }=\frac{d}{dr}\ln \Psi +\beta ^{\prime }.
\end{equation}%
Hence, Eq.~(\ref{modKGeq0}) can be reformulated as
\bea\label{KG1}
-\left[ \frac{\phi ^{\prime \prime }}{\phi ^{\prime }}-\frac{\phi ^{\prime }%
}{2\phi }+\frac{2}{r}+\frac{1}{2}\frac{d}{dr}\ln \Psi +\frac{1}{2}\beta
^{\prime }\right] e^{-\lambda }+\frac{\lambda ^{\prime }}{2}e^{-\lambda }
	\nonumber\\
+\frac{1}{3}\frac{\phi }{\phi ^{\prime }}\left[ 2V(\phi )-(1+\phi )V_{\phi
}(\phi )\right] =0.
\eea

We determine the function $\Psi $ by imposing the condition
\begin{equation}
\frac{d}{dr}\ln \Psi =-\frac{2\phi ^{\prime \prime }}{\phi ^{\prime }}+\frac{%
\phi ^{\prime }}{\phi }-\frac{4}{r},
\end{equation}%
which gives
\begin{equation}
\Psi =\Psi _{0}\frac{\phi }{r^{4}\phi ^{\prime 2}},
\end{equation}%
where $\Psi _{0}$ is an arbitrary constant of integration. Therefore, from Eq. (%
\ref{KG1}) we obtain
\begin{equation}
e^{-\lambda }\beta ^{\prime }=e^{-\lambda }\lambda ^{\prime }+U\left( \phi
\right) ,  \label{rel1}
\end{equation}%
where we have denoted
\begin{equation}
U\left( \phi \right) =\frac{2}{3}\frac{\phi }{\phi ^{\prime }}\left[ 2V(\phi
)-(1+\phi )V_{\phi }(\phi )\right] .
\end{equation}

From Eq. (\ref{rel1}), we immediately obtain
\begin{equation}
\beta =\lambda +\int e^{\lambda \left( r^{\prime }\right) }U\left( \phi
\left( r^{\prime }\right) \right) dr^{\prime }+C,
\end{equation}%
where $C$ is an arbitrary constant of integration, and
\begin{equation}\label{68}
\nu =\lambda +\ln \frac{\Psi _{0}\phi }{r^{4}\phi ^{\prime 2}}+\int
e^{\lambda \left( r^{\prime }\right) }U\left( \phi \left( r^{\prime }\right)
\right) dr^{\prime }+C.
\end{equation}

From the generalized Klein-Gordon  Eq. (\ref{modKGeq0}) we can express the term $\phi ^{\prime }\lambda
^{\prime }e^{-\lambda }/2\left( 1+\phi \right) $ as
\begin{equation}
\frac{\phi ^{\prime }\lambda ^{\prime }e^{-\lambda }}{2\left( 1+\phi \right)
}=\frac{\phi ^{\prime }e^{-\lambda }}{1+\phi }\left( \frac{\phi ^{\prime
\prime }}{\phi ^{\prime }}+\frac{\nu ^{\prime }}{2}-\frac{\phi ^{\prime }}{%
2\phi }+\frac{2}{r}\right) -\frac{1}{2}\frac{\phi ^{\prime }U\left( \phi
\right) }{1+\phi }.
\end{equation}

After substitution in Eq. ~(\ref{rho}) we obtain
\begin{eqnarray}
\frac{1}{r^{2}}\left( 1-e^{-\lambda }\right) +\frac{\lambda ^{\prime
}e^{-\lambda }}{r}+\frac{e^{-\lambda }\phi ^{\prime }}{1+\phi }\left( \frac{%
\nu ^{\prime }}{2}+\frac{1}{4}\frac{\phi ^{\prime }}{\phi }\right)
	\nonumber\\
-\frac{1}{%
2}\frac{\phi ^{\prime }U\left( \phi \right) }{1+\phi }-\frac{V\left( \phi
\right) }{2\left( 1+\phi \right) }=0.
\end{eqnarray}

Then, after subtracting Eqs.~(\ref{rho}) and (\ref{hybrid_pr}), and with the
use of the relation $\left( \nu ^{\prime }-\lambda ^{\prime }\right)
e^{-\lambda }/r=\left( e^{-\lambda }/r\right) d\ln \Psi /dr+U\left( \phi
\right) /r$, we obtain the equation
\begin{eqnarray}
\frac{1}{r}e^{-\lambda }\frac{d}{dr}\ln \Psi -\frac{2}{r^{2}}\left(
1-e^{-\lambda }\right) +\frac{e^{-\lambda }\phi ^{\prime }}{1+\phi }\frac{d}{%
dr}\ln r^{2}\sqrt{\phi }
	\nonumber\\
+\frac{U\left( \phi \right) }{2}\left( 1-\frac{1}{2}%
\frac{r\phi ^{\prime }}{1+\phi }\right) -\frac{V\left( \phi \right) }{%
2\left( 1+\phi \right) }=0,
\end{eqnarray}%
which allows us to obtain $e^{-\lambda }$ as
\begin{equation}
e^{-\lambda }=\frac{1+\frac{V\left( \phi \right) r^{2}}{2\left( 1+\phi
\right) }-\frac{U\left( \phi \right) r}{2}\left( 1+\frac{1}{2}\frac{r\phi
^{\prime }}{1+\phi }\right) }{1+\frac{r}{2}\frac{d}{dr}\ln \Psi +\frac{1}{2}%
\frac{\phi ^{\prime }r^{2}}{1+\phi }\frac{d}{dr}\ln r^{2}\sqrt{\phi }}.
\end{equation}

For the effective mass function $M_{eff}=\left( c^{2}/2G\right) r\left(
1-e^{-\lambda }\right) $ we obtain
\begin{widetext}
\begin{equation}
M_{eff}=\frac{c^{2}}{2G}r^{2}\frac{\frac{1}{2}\frac{d}{dr}\ln \Psi +\frac{1}{%
2}\frac{\phi ^{\prime }r}{1+\phi }\frac{d}{dr}\ln r^{2}\sqrt{\phi }-\frac{%
V\left( \phi \right) r}{2\left( 1+\phi \right) }+\frac{U\left( \phi \right)
}{2}\left( 1+\frac{1}{2}\frac{r\phi ^{\prime }}{1+\phi }\right) }{1+\frac{r}{%
2}\frac{d}{dr}\ln \Psi +\frac{1}{2}\frac{\phi ^{\prime }r^{2}}{1+\phi }\frac{%
d}{dr}\ln r^{2}\sqrt{\phi }}.
\end{equation}
\end{widetext}

Once the metric tensor component $e^{-\lambda}$ is known, the metric tensor component $e^{\nu}$ can be obtained from Eq.~(\ref{68}). In the case $V\left( \phi (r)\right) \equiv 0$, the above equations take the
form
\begin{equation}
e^{-\lambda }=\frac{1}{1+\frac{r}{2}\frac{d}{dr}\ln \Psi +\frac{1}{2}\frac{%
\phi ^{\prime }r^{2}}{1+\phi }\frac{d}{dr}\ln r^{2}\sqrt{\phi }},
\end{equation}
and
\begin{equation}
M_{eff}=\frac{c^{2}}{2G}r^{2}\frac{\frac{1}{2}\frac{d}{dr}\ln \Psi +\frac{1}{2}\frac{%
\phi ^{\prime }r}{1+\phi }\frac{d}{dr}\ln r^{2}\sqrt{\phi }}{1+\frac{r}{2}%
\frac{d}{dr}\ln \Psi +\frac{1}{2}\frac{\phi ^{\prime }r^{2}}{1+\phi }\frac{d%
}{dr}\ln r^{2}\sqrt{\phi }},
\end{equation}%
respectively. Hence in the HMPG theory the geometric as well as the physical
properties of the gravitational field in the vacuum are completely
determined by the scalar field $\phi $, and of its derivatives.

\section{Numerical black hole solutions of the vacuum field equations in HMPG}\label{sect4}

Since the system of equations describing the vacuum static spherically symmetric
gravitational field does not seem to admit any simple exact analytical
solution of Schwarzschild or de Sitter type, in the following we will concentrate on the numerical solutions of
the system of Eqs. (\ref{53})-(\ref{57}), with the initial condition given
by Eqs. (\ref{58}). These equations are formulated in the variable $\xi =1/r$%
, and to obtain their solutions we start the integration at $\xi =\xi
_{\infty}$, corresponding to a very large distance from the central object, i.e., spatial infinity, and to very small values of $\xi $. The presence of
the singularity, and of the black hole horizon, is detected as the zeros of
the metric tensor coefficients $e^{\nu}$ and $e^{-\lambda}$, respectively.
In our analysis we consider several forms of the potential $V$ of the scalar
field.

\subsection{The case $V(\phi)=0$}

\begin{figure*}[!htb]
\centering
\includegraphics[scale=0.65]{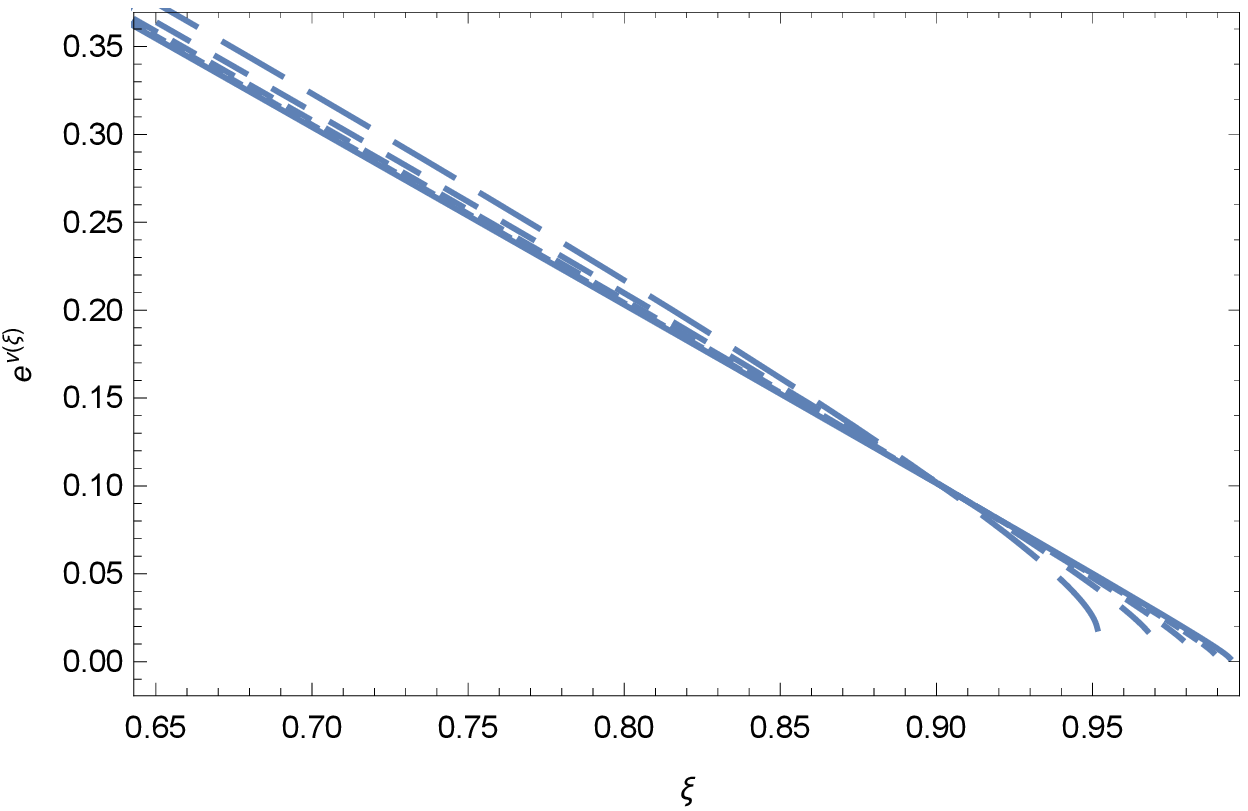} %
\includegraphics[scale=0.65]{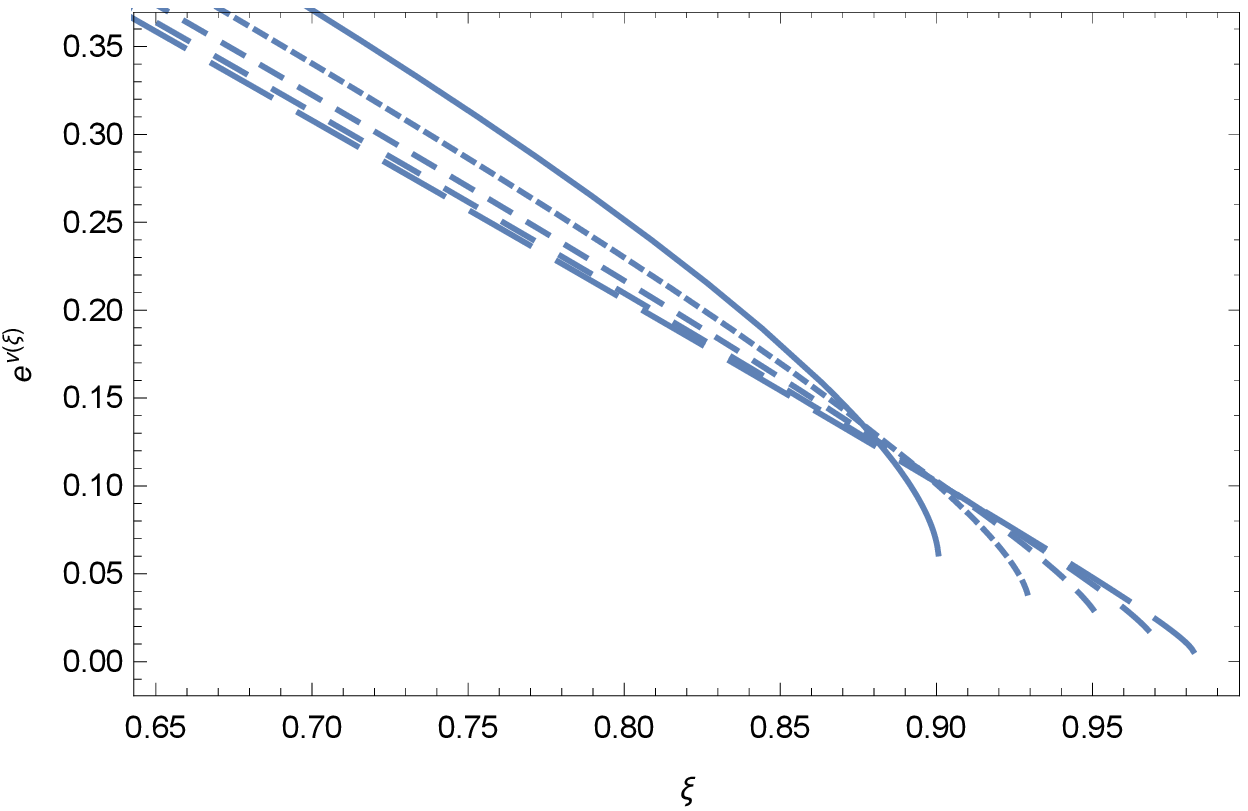}
\caption{Variation of the metric tensor components $e^{\nu}$ in the vacuum outside a spherically symmetric compact
object in HMPG with a vanishing scalar field potential. Left figure: The initial value of scalar field, $\phi_0$, is fixed at $\phi_0 = 1$ while the initial value of its derivative is taken to be: $u_0=4 \times 10^{-9}$ (solid curve), $u_0=8 \times 10^{-9}$ (dotted curve), $u_0=1.6 \times 10^{-8}$ (short dashed curve), $u_0=3.2 \times 10^{-8}$ (dashed curve), $u_0=6.4 \times 10^{-8}$ (long dashed curve). Right figure: The initial value of the derivative of the scalar field, $u_0$, is fixed at $u_0=5.12\times 10^{-7}$ while the initial value of the scalar field is taken to be: $\phi_0=0.5$ (solid curve), $\phi_0=1$ (dotted curve), $\phi_0=2$ (short dashed curve), $\phi_0=4$ (dashed curve), $\phi_0=8$ (long dashed curve).}
\label{fig1}
\end{figure*}

As a first example of numerical vacuum solutions in HMPG, we consider the case $V(\phi)=0$. In order to numerically integrate the gravitational field equations Eqs. (\ref{53})-(\ref{57}) we need to fix the initial values of the scalar field $\phi$, and of its derivative at infinity, corresponding to the value $\xi =0$ of the dimensionless radial coordinate $\xi$. As for the metric we assume that at infinity it is Minkowskian. Hence the nature of the central singularity in HMPG is essentially determined by the numerical values  the field $\phi $ and its derivative $\phi '$ takes at infinity. In order to investigate the effect of the initial conditions we consider two different classes of solutions. In the first class we assume that the initial value of the field at infinity is fixed, and we let its derivatives vary. For the second set of models we take the derivative of the scalar field as fixed at infinity, and we investigate the effects of the field variation on the geometry. The variations of the metric tensor coefficient $e^{\nu}$ for these two cases is presented in Figs.~\ref{fig1}.

\begin{figure*}[!htb]
\centering
\includegraphics[scale=0.65]{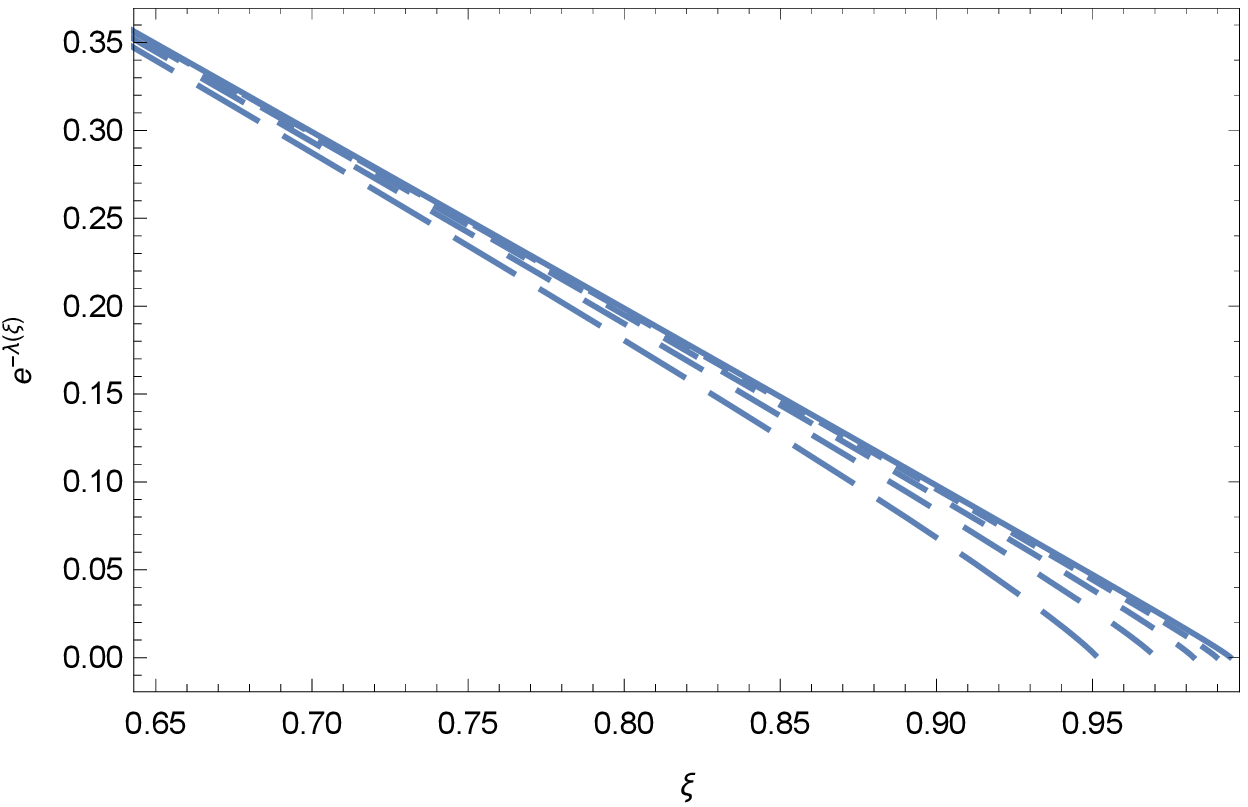} %
\includegraphics[scale=0.65]{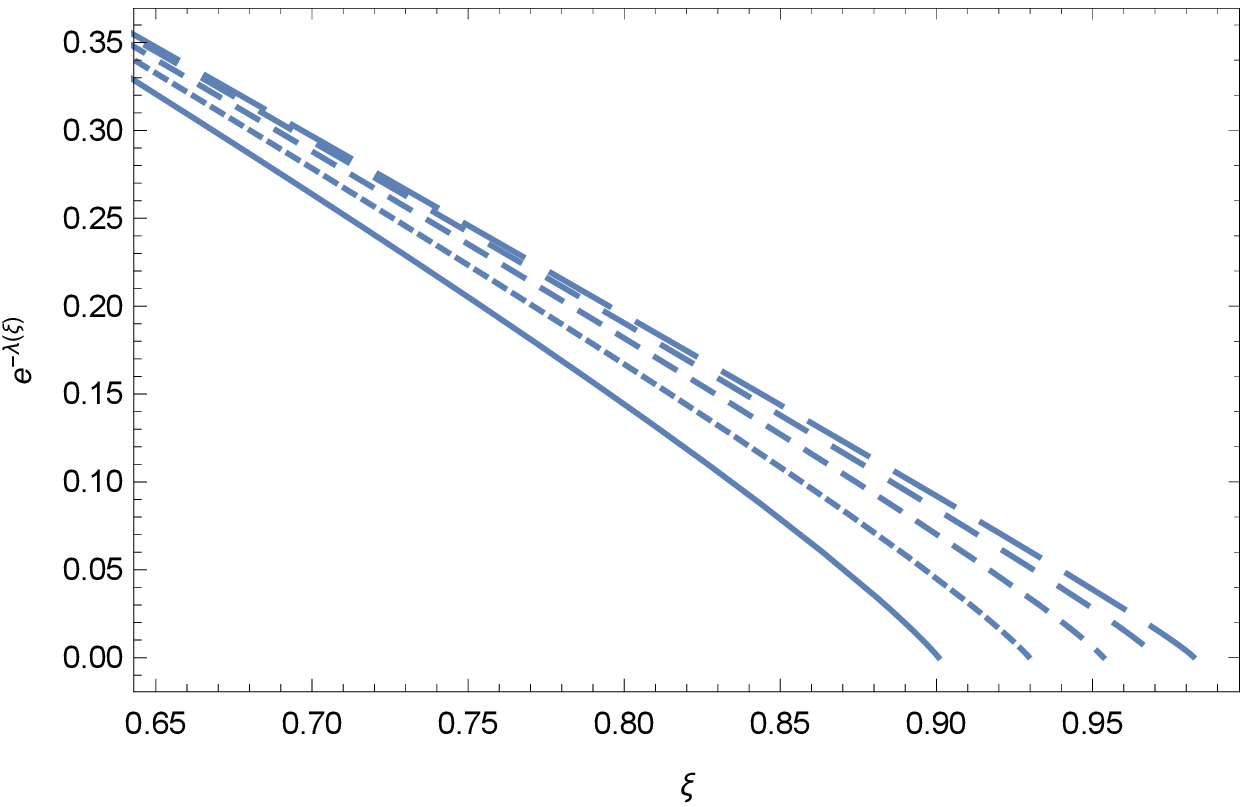}
\caption{Variation of the metric tensor components $e^{-\lambda}$ in the vacuum outside a spherically symmetric compact
object in HMPG with a vanishing scalar field
potential. Left: The initial value of scalar field, $\phi_0$, is fixed at $\phi_0 = 1$ while the initial value of its derivative is taken to be: $u_0=4 \times 10^{-9}$ (solid curve), $u_0=8 \times 10^{-9}$ (dotted curve), $u_0=1.6 \times 10^{-8}$ (short dashed curve), $u_0=3.2 \times 10^{-8}$ (dashed curve), $u_0=6.4 \times 10^{-8}$ (long dashed curve). Right: The initial value of the derivative of the scalar field, $u_0$, is fixed at $u_0=5.12\times 10^{-7}$ while the initial value of the scalar field is taken to be: $\phi_0=0.5$ (solid curve), $\phi_0=1$ (dotted curve), $\phi_0=2$ (short dashed curve), $\phi_0=4$ (dashed curve), $\phi_0=8$ (long dashed curve).}
\label{fig2}
\end{figure*}

 As one can see from the figures describing the variation of $e^{\nu}$, at fixed values of $\xi =\xi_S$ the metric becomes singular. The same effect can be also observed in the case of the evolution of $e^{-\lambda}$, presented in Figs.~\ref{fig2}. For both metric tensor coefficients a singular behavior does appear for a finite value of $\xi$, indicating the formation of an outer
apparent horizon, and of a  black hole. However, the position of the outer
apparent horizon $\xi _S$ covering the black hole depends on the initial values at infinity of the scalar field.

\begin{figure*}[!htb]
\centering
\includegraphics[scale=0.65]{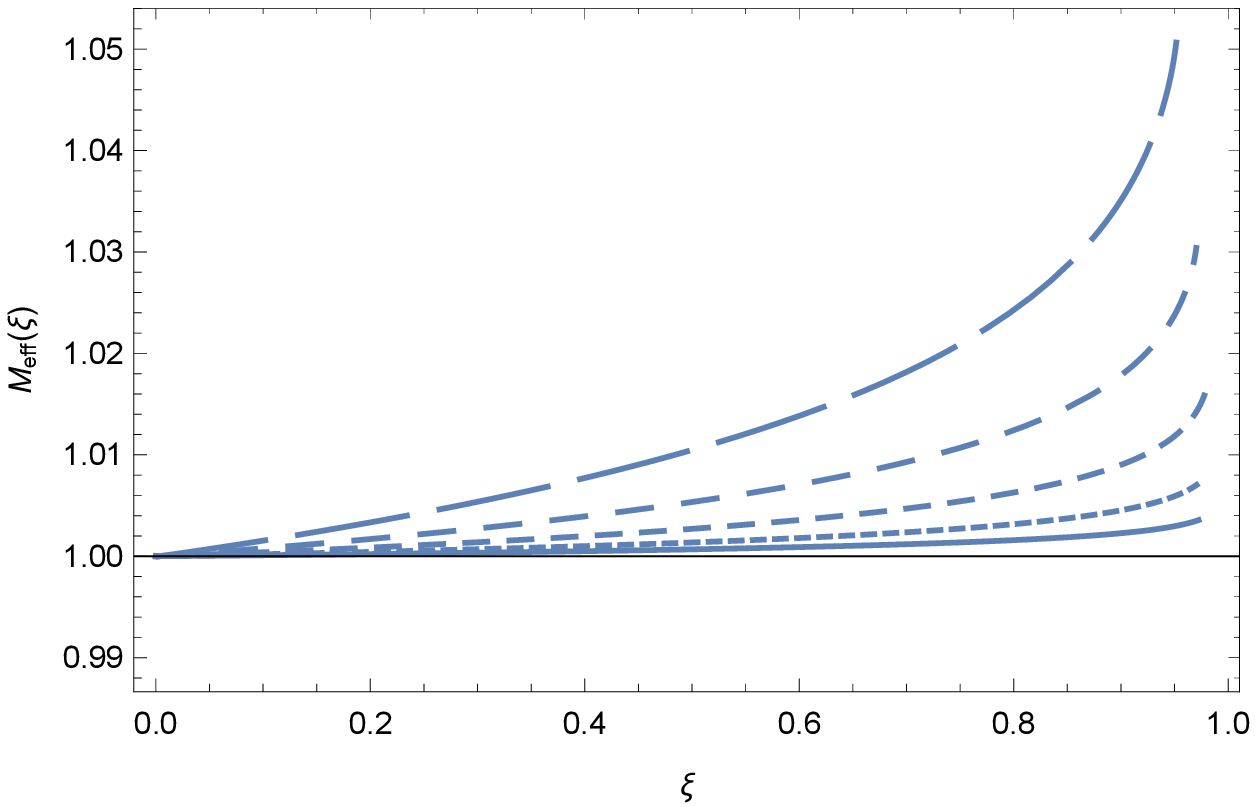}
\includegraphics[scale=0.65]{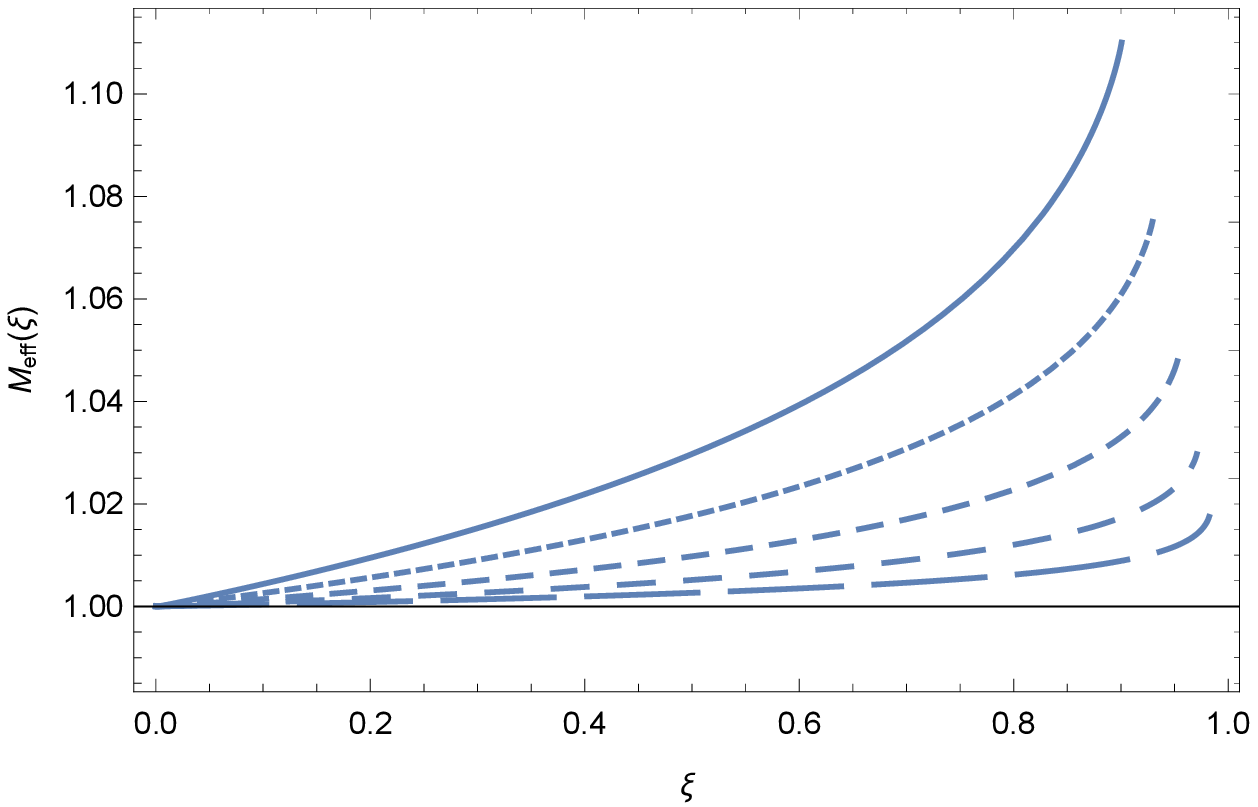}
\caption{Variation of the effective mass function $M_{eff}(\xi)$ in the vacuum
outside a spherically symmetric compact object in HMPG with a vanishing scalar field potential. Left: The initial value of scalar field, $\phi_0$, is fixed at $\phi_0 = 1$ while the initial value of its derivative is taken to be: $u_0=4 \times 10^{-9}$ (solid curve), $u_0=8 \times 10^{-9}$ (dotted curve), $u_0=1.6 \times 10^{-8}$ (short dashed curve), $u_0=3.2 \times 10^{-8}$ (dashed curve), $u_0=6.4 \times 10^{-8}$ (long dashed curve). Right: The initial value of the derivative of the scalar field, $u_0$, is fixed at $u_0=5.12\times 10^{-7}$ while the initial value of the scalar field is taken to be: $\phi_0=0.5$ (solid curve), $\phi_0=1$ (dotted curve), $\phi_0=2$ (short dashed curve), $\phi_0=4$ (dashed curve), $\phi_0=8$ (long dashed curve).}
\label{fig3}
\end{figure*}

The variation of the effective mass of the black hole is represented in Figs.~\ref{fig3}. As indicated by the figures, the mass of the black hole significantly increases as compared to its mass at infinity, where the effects of the scalar field are neglected. Hence, in HMPG the scalar field gives a significant contribution to the mass of the gravitating object.

The exact locations of the outer
apparent horizon for HPMG with a vanishing scalar field potential are represented in Table~\ref{table1} and \ref{table2}, respectively.
\begin{table}
\begin{center}
\begin{tabular}{| c|c|}
\hline
$u_0$ &  $\xi_S$ \\
\hline
$5\times 10^{-10}$ & 0.99856 \\
\hline
$1\times 10^{-9}$ & 0.99781 \\
\hline
$2\times 10^{-9}$ & 0.99641 \\
\hline
$4\times 10^{-9}$ & 0.99391 \\
\hline
$8\times 10^{-9}$ & 0.98956 \\
\hline
$1.6\times 10^{-8}$ & 0.98219 \\
\hline
$3.2\times 10^{-8}$ & 0.97020 \\
\hline
$6.4\times 10^{-8}$ & 0.95155 \\
\hline
$1.28\times 10^{-7}$ & 0.92393 \\
\hline
$2.56\times 10^{-7}$ & 0.88450 \\
\hline
$5.12\times 10^{-7}$ & 0.82693 \\
\hline
\end{tabular}
\caption{Values of $\xi=\xi _S$ where the singularity occurs, corresponding to the radius of the outer apparent horizon, for fixed $\phi_0=1$ and different values of $\phi '(0)=u_0$ for the case of the vanishing scalar field potential $V=0$.}\label{table1}
\end{center}
\end{table}
\begin{table}
\begin{center}
\begin{tabular}{| c|c|}
\hline
$\phi_0$ &  $\xi _S$ \\
\hline
0.5 & 0.77061 \\
\hline
1 & 0.82693 \\
\hline
2 & 0.86764 \\
\hline
4 & 0.90060 \\
\hline
8 & 0.92965 \\
\hline
16 & 0.95335 \\
\hline
32 & 0.97065 \\
\hline
64 & 0.98219 \\
\hline
128 & 0.98939 \\
\hline
256 &0.99366 \\
\hline
512 & 0.99610 \\
\hline
\end{tabular}
\caption{Values of $\xi=\xi_S$ where the singularity in the static vacuum field equations of HMPG occurs, for fixed $u_0=5.12\times 10^{-7}$ and for different values of $\phi _0$ in the case of the vanishing scalar field potential $V=0$.}\label{table2}
\end{center}
\end{table}
As one can see from Table~\ref{table1}, for very small values of $u_0$, the position of the outer
apparent horizon of the HMPG black hole almost coincides with the Schwarzschild radius of the black hole $\xi _S=1$. With the increase of $u_0$ there is a significant decrease in the numerical values of the metric singularity, which can reach values as low as 0.8 of the Schwarzschild radius, indicating that the presence of the scalar field pushes the outer
apparent horizon towards the center of the black hole. A different trend can be observed from the numerical results presented in Table~\ref{table2}. For a fixed but small $u_0$, the position of the outer
apparent horizon is inversely proportional to the initial values of the scalar field. The outer
apparent horizon of the black hole approaches the Schwarzschild radius for large initial values of the field, while small values of $\phi _0$ of the order of one lead to a significant decrease in the position of $\xi _S$.

\subsubsection{Fitting of the numerical results}

\begin{figure}[!htb]
\centering
\includegraphics[scale=0.65]{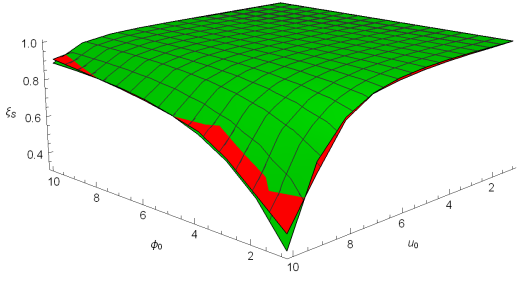}
\caption{Comparison of the fitting function $\xi _S\left(\phi_0,u_0\right)=1 - 148411 \times  u_0/ \phi_0 - 2.737 \times 10^{11} \times u_0^2$ for the position of the event horizon of the HMPG theory black holes and the numerical data for the vanishing potential case, for $\phi_0 \in [0.15, 5.7665]$ and $u_0 \in [3\times 10^{-11}, 5.9\times 10^{-7}]$. }
\label{fig7aa}
\end{figure}

As a function of the initial conditions for the scalar field the expression of the outer
apparent horizon of the black hole can be obtained as
\be
\xi _S\left(\phi_0,u_0\right)=1 - 148411 \times \frac{ u_0}{ \phi_0} - 2.737 \times 10^{11} \times u_0^2,
\ee
with an $R$ squared value of $R^2=0.99926$. The comparison of the numerical results and of the fitting function is presented in Fig.~\ref{fig7aa}.

We have also obtained numerical fits for the three functions, $M_{eff}(\xi)$, $e^{\nu(\xi)}$ and $e^{-\lambda(\xi)}$, for the same combination of parameters $\phi_0, u_0$ as in Table \ref{table1} and Table \ref{table2}.

For the mass function we consider a representation of the form
\begin{equation}\label{mass1}
M_{eff}(\xi)= A_M + B_M \xi + C_M \xi^2,
\end{equation}
where we have further obtained the coefficients $A_M$, $B_M$, $C_M$ as functions $A_M=A_M\left(\phi_0, u_0\right)$, $B_M=B_M\left(\phi_0,u_0\right)$, and $C_M=C_M\left(\phi_0,u_0\right)$ as given below,
\begin{eqnarray}
A_M&=&1.00067 + 53919.2 u_0 - 4814.28 \times \phi_0 u_0
	\nonumber\\
&&-3.34854\times 10^{10}\times  u_0^2-
\frac{2.72635\times 10^9 u_0^2}{\phi_0^2}
	\nonumber\\
&&+ \frac{6.44955\times 10^9 \times u_0^2}{\phi_0},
\end{eqnarray}
with an $R$ squared of $R^2=0.999998$,
\begin{eqnarray}
B_M&=&-213635 u_0 + \frac{23702.9 u_0}{\phi_0} + 18747.6 \times \phi_0 u_0 \nonumber\\
&&+ 1.52191\times 10^{11} u_0^2
 + \frac{3.86999\times 10^{10} u_0^2}{\phi_0^2}
 	\nonumber\\
 &&- \frac{5.94209\times 10^{10} u_0^2}{\phi_0},
\end{eqnarray}
with with $R^2=0.965492$, and finally
\begin{eqnarray}
C_M&=&639111 u_0 + \frac{25555.1 u_0}{\phi_0} - 39078.4 \phi_0 u_0 \nonumber\\
&&- 5.82573\times  10^{11} u_0^2
- \frac{5.7583\times  10^{10} u_0^2}{\phi_0^2}
	\nonumber\\
&&+\frac{5.73072\times 10^{11} u_0^2}{\phi_0},
\end{eqnarray}
with an $R$ squared of $R^2=0.990561$, respectively.

For the metric tensor coefficient $e^{\nu(\xi)}$, we consider a representation of the form
\begin{equation}
e^{\nu(\xi)}= A_{\nu} + B_{\nu} \xi + C_{\nu} \xi^2,
\end{equation}
where  the coefficients $A_{\nu}$, $B_{\nu}$, and $C_{\nu}$ are given as functions of the initial values of the scalar field as
\begin{eqnarray}
A_{\nu}&=&1.01476 - 165700. u_0 + 17233.5 \phi_0 u_0
	\nonumber\\
&&+9.14106\times 10^{10} u_0^2 +
\frac{7.59538\times 10^9 u_0^2}{\phi_0^2}
	\nonumber\\
&&- \frac{3.95859\times 10^{10} u_0^2}{\phi_0},
\end{eqnarray}
with an $R$ squared of $R^2=0.999985$,
\begin{eqnarray}
B_{\nu}&=&-0.995931 + 2.28588\times 10^6 u_0 - 231896 \phi_0 u_0
	\nonumber\\
&&- 1.36511\times 10^{12} u_0^2 - \frac{1.13951\times 10^{11} u_0^2}{\phi_0^2}
	\nonumber\\
&&+\frac{1.01312\times 10^{12} u_0^2}{\phi_0},
\end{eqnarray}
with $R^2=0.997263$, and finally
\begin{eqnarray}
C_{\nu}&=&-1.54598\times 10^6 u_0 - \frac{255308 u_0}{\phi_0}
	\nn \\
&&	+120339 \phi_0 u_0 +
 1.18476\times 10^{12} u_0^2
 	\nn \\
 &&	+\frac{1.24333\times 10^{11} u_0^2}{\phi_0^2} - \frac{1.6744\times 10^{12} u_0^2}{\phi_0},
\end{eqnarray}
with an $R$ squared of $R^2=0.997204$.

For the metric tensor component $e^{-\lambda(\xi)}$, we adopt the functional form
\begin{eqnarray}
e^{-\lambda(\xi)}&=&1-\xi M_{eff}(\xi)=1-(A_M + B_M \xi + C_M \xi^2)
	\nonumber\\
&=&1+A_{\lambda}\xi +B_{\lambda}\xi^2+C_{\lambda}\xi^3,
\end{eqnarray}
where the coefficients $A_{\lambda}$, $B_{\lambda}$, $C_{\lambda}$ are given as functions of the initial conditions at infinity of the scalar field as
\begin{eqnarray}
A_{\lambda}&=&-0.933912 - 108822 u_0 + 10973 \phi_0 u_0
	\nonumber\\
&&+4.6961\times 10^{10} u_0^2 +\frac{6.11265\times 10^9 u_0^2}{\phi_0^2}
	\nonumber\\
&&+ \frac{1.81929\times 10^{10} u_0^2}{\phi_0},
\end{eqnarray}
with an $R$ squared of $R^2=0.999976$,
\begin{eqnarray}
B_{\lambda}&=&-0.121357 + 357381 u_0 - 41431 \phi_0 u_0
	\nonumber\\
&&-1.12908\times 10^{11} u_0^2-
\frac{7.12164\times 10^{10} u_0^2}{\phi_0^2}
	\nonumber\\
&&- \frac{8.9032\times 10^9 u_0^2}{\phi_0},
\end{eqnarray}
with $R^2=0.981236$, and
\begin{eqnarray}
C_{\lambda}&=&0.0645491 - 777792 u_0 - \frac{36125 u_0}{\phi_0}
	\nonumber\\
&&+ 59800.9 \phi_0 u_0+ 6.11368\times 10^{11} u_0^2
	\nonumber\\
&&+ \frac{1.16131\times 10^{11} u_0^2}{\phi_0^2}-
 \frac{6.21693\times 10^{11} u_0^2}{\phi_0},
\end{eqnarray}
with an $R$ squared of $R^2=0.961695$, respectively.

\subsection{The Higgs-type potential: $V(\phi)=-\frac{\mu ^2}{2}\phi ^2+\frac{\varsigma}{4}\phi ^4$}

\begin{figure*}[!htb]
\centering
\includegraphics[scale=0.65]{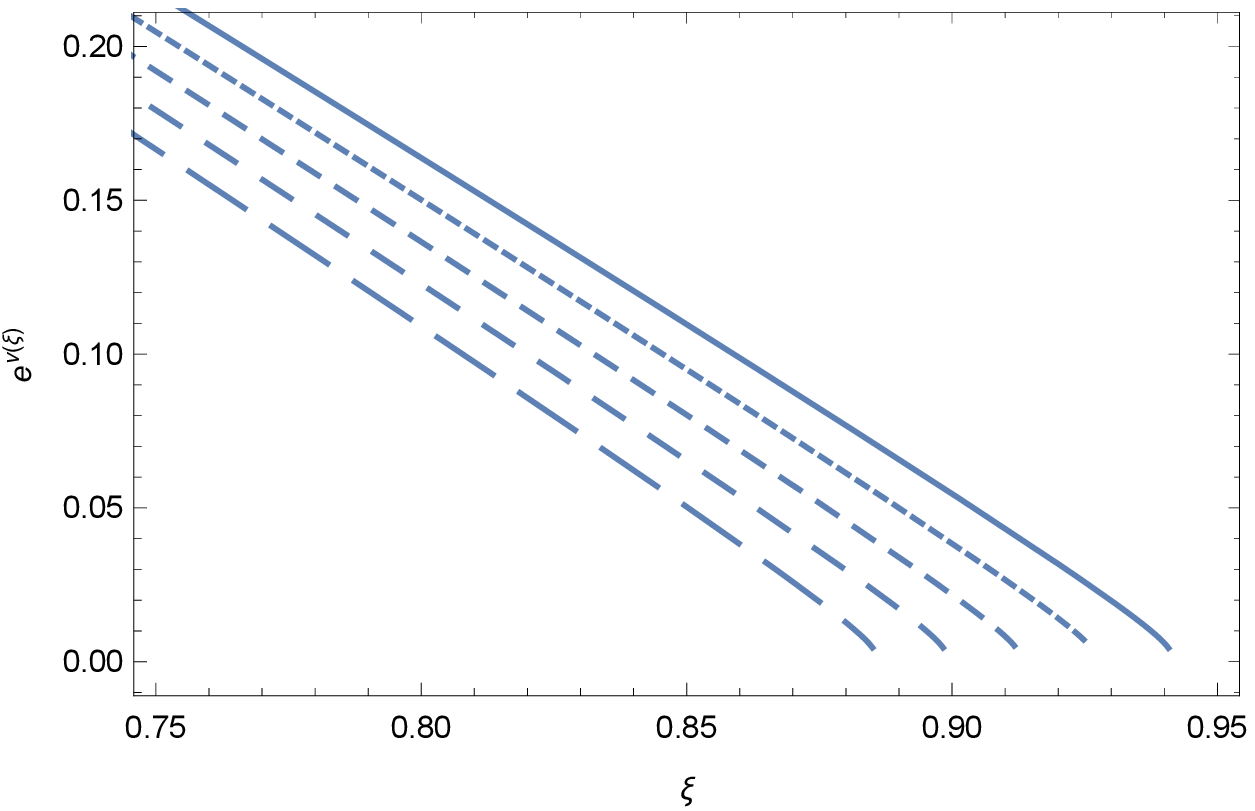} %
\includegraphics[scale=0.65]{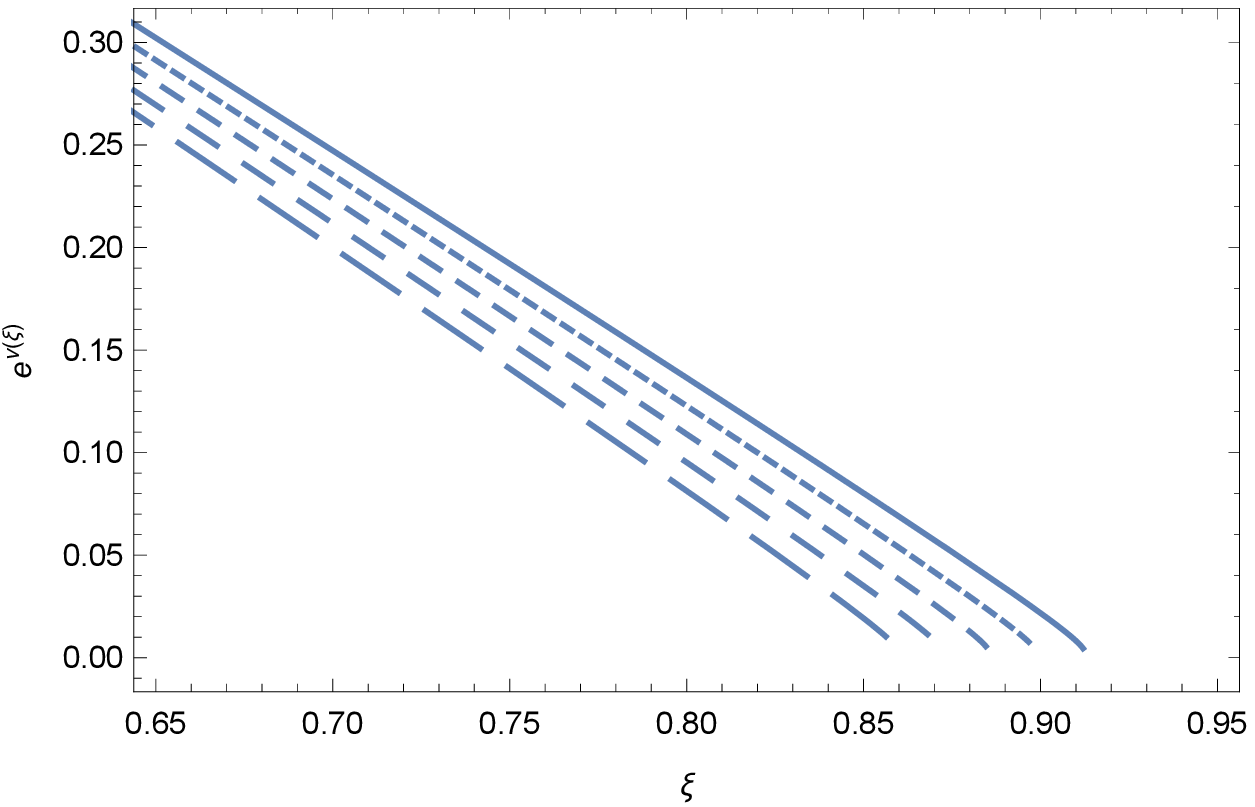}
\caption{Variation of the metric tensor components $e^{\nu}$  in the vacuum outside a spherically symmetric compact
object in HMPG with a Higgs-type potential $V=\alpha \phi^2+\beta \phi ^4$ of the scalar field. The initial value of the scalar field is fixed at $\phi_0=1$, and its derivative is fixed at $u_0=10^{-8}$, respectively. Left figure: the parameter $\alpha$ is fixed at $\alpha=10^{-10}$ while the parameter $\beta$ is taken to be: $\beta=2\times 10^{-10}$ (solid curve), $\beta=3\times 10^{-10}$ (dotted curve), $\beta=4\times 10^{-10}$ (short dashed curve), $\beta=5\times 10^{-10}$ (dashed curve), $\beta=6\times 10^{-10}$ (long dashed curve). Right figure: the parameter $\beta$ is fixed at $\beta=10^{-10}$ while the parameter $\alpha$ is taken to be: $\alpha=2\times 10^{-10}$ (solid curve), $\alpha=3\times 10^{-10}$ (dotted curve), $\alpha=4\times 10^{-10}$ (short dashed curve), $\alpha=5\times 10^{-10}$ (dashed curve), $\alpha=6\times 10^{-10}$ (long dashed curve).}
\label{fig4}
\end{figure*}

\begin{figure*}[!htb]
\centering
\includegraphics[scale=0.65]{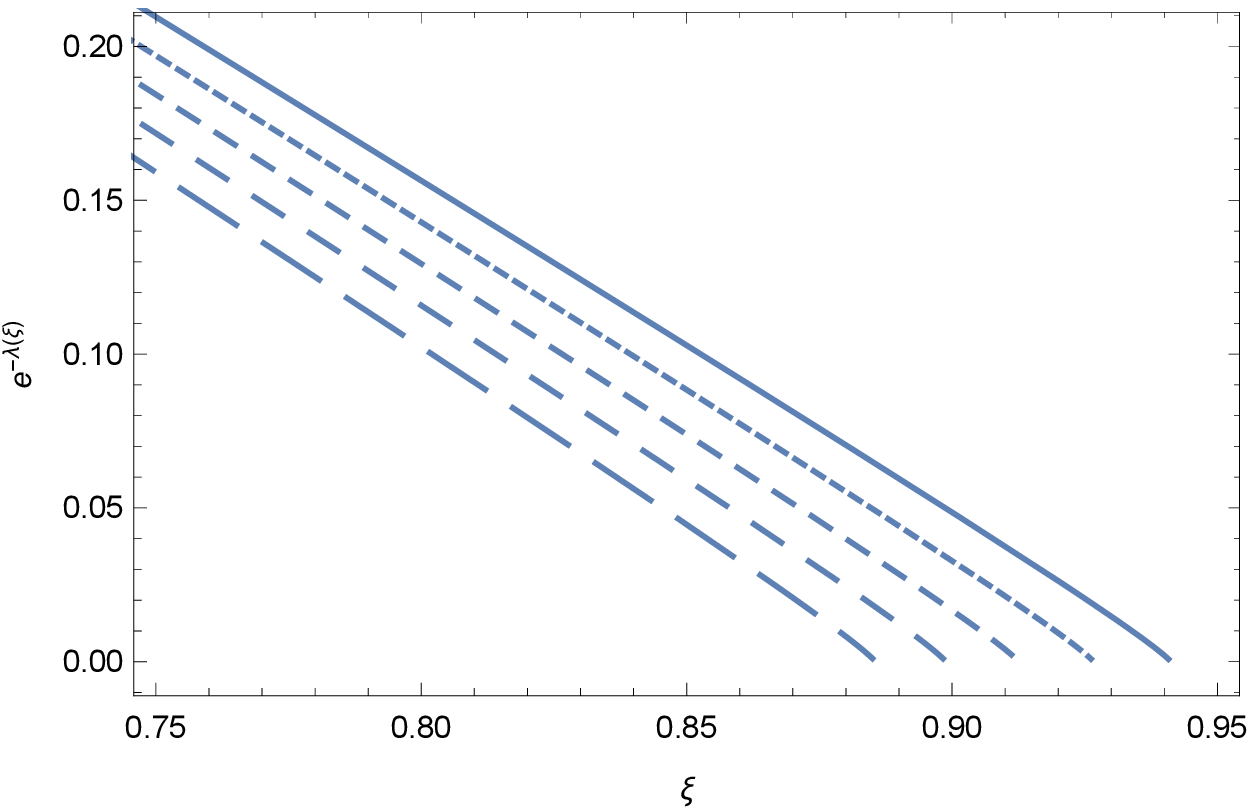} %
\includegraphics[scale=0.65]{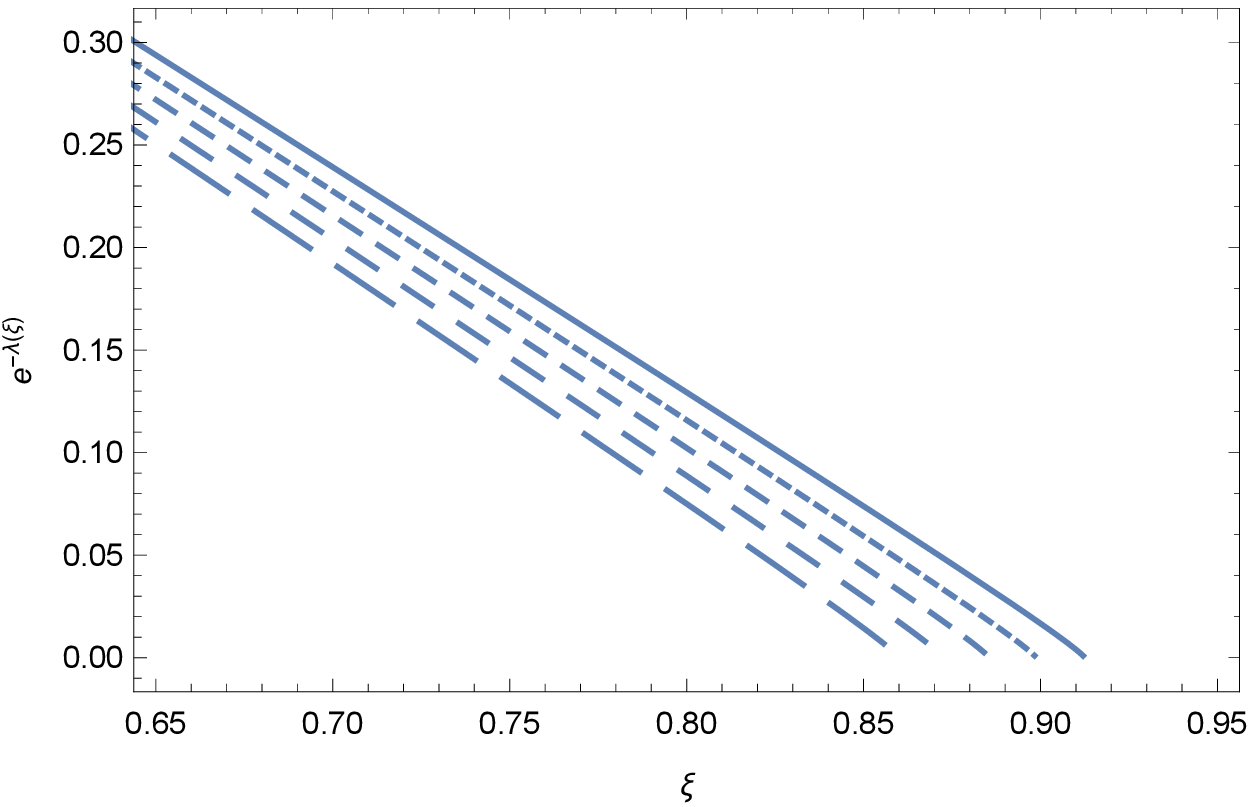}
\caption{Variation of the metric tensor components $e^{\lambda}$ in the vacuum outside a spherically symmetric compact
object in HMPG with a Higgs-type potential $V=\alpha \phi^2+\beta \phi ^4$ of the scalar field. The initial value of the scalar field is fixed at $\phi_0=1$, and its derivative is fixed at $u_0=10^{-8}$, respectively. Left figure: the parameter $\alpha$ is fixed at $\alpha=10^{-10}$ while the parameter $\beta$ is taken to be: $\beta=2\times 10^{-10}$ (solid curve), $\beta=3\times 10^{-10}$ (dotted curve), $\beta=4\times 10^{-10}$ (short dashed curve), $\beta=5\times 10^{-10}$ (dashed curve), $\beta=6\times 10^{-10}$ (long dashed curve). Right figure: the parameter $\beta$ is fixed at $\beta=10^{-10}$ while the parameter $\alpha$ is taken to be: $\alpha=2\times 10^{-10}$ (solid curve), $\alpha=3\times 10^{-10}$ (dotted curve), $\alpha=4\times 10^{-10}$ (short dashed curve), $\alpha=5\times 10^{-10}$ (dashed curve), $\alpha=6\times 10^{-10}$ (long dashed curve).}
\label{fig5}
\end{figure*}

As another example of vacuum solutions of the gravitational field equations
in HMPG we consider the case of the scalar field with Higgs-type potential,
\be
V(\phi)=-\frac{\mu ^2}{2}\phi ^2+\frac{\varsigma}{4}\phi ^4,
\ee
where $\mu ^2$ and $\varsigma $ are constants. The Higgs potential plays a fundamental role in particle physics, and by analogy with quantum field theoretical models we assume that $-\mu ^2$ gives the mass of the scalar field particle associated to HMPG. The Higgs self-coupling constant $\varsigma$ takes the value $\varsigma \approx 1/8$ for the case of strong interactions \cite{Higgs}. This value
is obtained from the determination of the mass of the Higgs boson from accelerator experiments, but the self-interacting properties of the scalar field in HMPG may be very different than those suggested by QCD. By taking into account the new variable introduced in the present approach the scalar field potential can be written in a dimensionless form as
\be
v(\phi)=\alpha \phi ^2+\beta \phi ^4,
\ee
where
\be
\alpha =-\frac{1}{4}\left(\frac{2GnM_{\odot}}{c^2}\right)^2\mu ^2, \quad
\beta =\frac{1}{2}\left(\frac{2GnM_{\odot}}{c^2}\right)^2\varsigma.
\ee
The Higgs-type potential generates  four-parameter ($\alpha$, $\beta$,  $\phi _0$, $u_0$) classes of solutions of the static gravitational field equations in HMPG. However, in the following, we will restrict our analysis to the investigation of the role played by the constants $\alpha $ and $\beta $ of the potential in the formation of the event horizon of the black holes. Hence, we fix $\phi _0$ and $u_0$, and vary the numerical values of $\alpha $ and $\beta$.
The variations with respect to $\xi$ of the metric tensor components and of the
mass function are represented, for fixed values of $\phi _0$ and $u_0$  in Figs.~\ref{fig4} and \ref{fig5}, respectively.

Similarly to the case of the zero scalar field potential, the metric tensor components become singular at finite values  of the radial coordinate $\xi$, indicating the presence of an event horizon, and the formation of a black hole. The position of the event horizon strongly depends on the model parameters, with this dependence exemplified in Table~\ref{table3} and \ref{table4}, respectively.

\begin{table}
\begin{center}
\begin{tabular}{| c|c|}
\hline
$\beta$ &  $\xi_S$ \\
\hline
$2\times 10^{-10}$ & 0.97140 \\
\hline
$3\times 10^{-10}$ & 0.95582 \\
\hline
$4\times 10^{-10}$ & 0.94075 \\
\hline
$5\times 10^{-10}$ & 0.92615 \\
\hline
$6\times 10^{-10}$ & 0.91201 \\
\hline
$7\times 10^{-10}$ & 0.89930 \\
\hline
$8\times10^{-10}$ & 0.88499 \\
\hline
$9\times 10^{-10}$ & 0.87207 \\
\hline
\end{tabular}
\caption{Values of $\xi=\xi_S$ where the singularity in the field equations occur, indicating the formation of an outer
apparent horizon, for fixed $\phi_0=1$, $u_0=10^{-8}$, $\alpha=10^{-10}$ and varying $\beta$ in the case of the Higgs potential, $V=\alpha \phi^2+\beta \phi ^4$.}\label{table3}
\end{center}
\end{table}

\begin{table}
\begin{center}
\begin{tabular}{| c|c|}
\hline
$\alpha$ &  $\xi_S$ \\
\hline
$-2\times 10^{-10}$ & 0.94075 \\
\hline
$-3\times 10^{-10}$ & 0.91201 \\
\hline
$-4\times 10^{-10}$ & 0.92615 \\
\hline
$-5\times 10^{-10}$ & 0.89829 \\
\hline
$-6\times 10^{-10}$ & 0.88498 \\
\hline
$-7\times 10^{-10}$ & 0.87207 \\
\hline
$-8\times 10^{-10}$ & 0.85952 \\
\hline
$-9\times 10^{-10}$ & 0.84734 \\
\hline
\end{tabular}
\caption{Values of $\xi=\xi _S$ where the singularity in the field equations occur, indicating the formation of an outer
apparent horizon, for fixed $\phi_0=1$, $u_0=10^{-8}$, $\beta=10^{-10}$ and varying $\alpha$ in the case of the Higgs potential, $V=\alpha \phi^2+\beta \phi^4$.}\label{table4}
\end{center}
\end{table}

\begin{figure*}[!htb]
\centering
\includegraphics[scale=0.65]{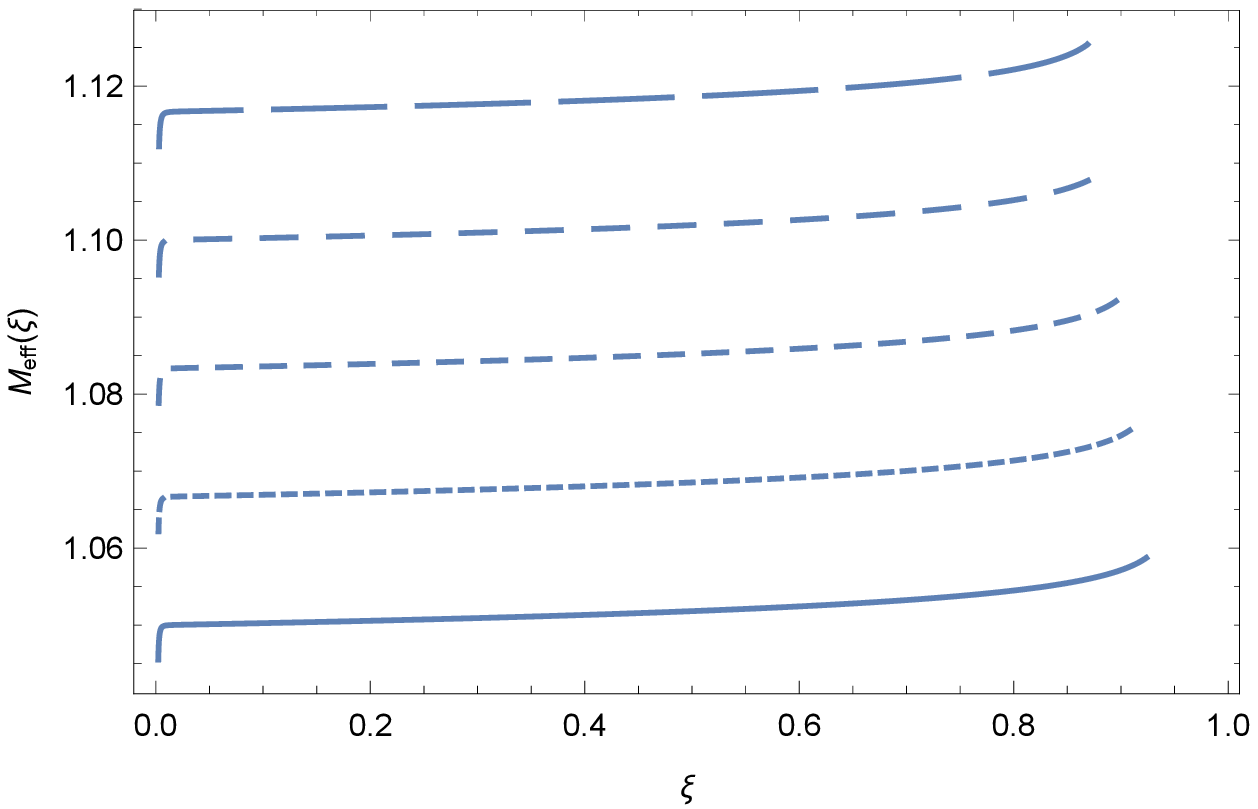}
\includegraphics[scale=0.65]{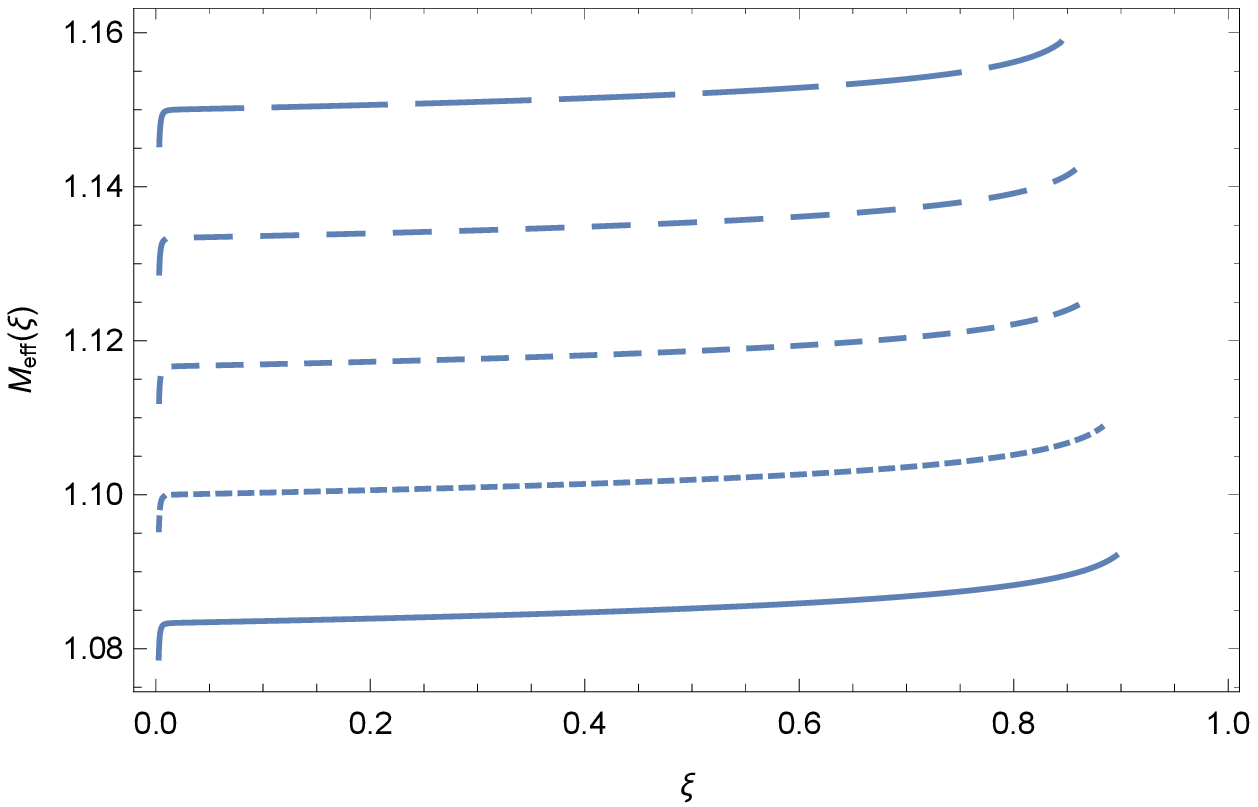}
\caption{Variation of the effective mass function $M_{eff}(\xi)$  in the vacuum outside a spherically symmetric compact
object in HMPG with a Higgs-type potential $V=\alpha \phi^2+\beta \phi ^4$ of the scalar field. The initial value of the scalar field is fixed at $\phi_0=1$, and its derivative is fixed at $u_0=10^{-8}$, respectively. Left figure: the parameter $\alpha$ is fixed at $\alpha=10^{-10}$ while the parameter $\beta$ is taken to be: $\beta=2\times 10^{-10}$ (solid curve), $\beta=3\times 10^{-10}$ (dotted curve), $\beta=4\times 10^{-10}$ (short dashed curve), $\beta=5\times 10^{-10}$ (dashed curve), $\beta=6\times 10^{-10}$ (long dashed curve). Right figure: the parameter $\beta$ is fixed at $\beta=10^{-10}$ while the parameter $\alpha$ is taken to be: $\alpha=2\times 10^{-10}$ (solid curve), $\alpha=3\times 10^{-10}$ (dotted curve), $\alpha=4\times 10^{-10}$ (short dashed curve), $\alpha=5\times 10^{-10}$ (dashed curve), $\alpha=6\times 10^{-10}$ (long dashed curve).}
\label{fig6}
\end{figure*}

The effective mass function $M(\xi)$, represented in Fig.~\ref{fig6}, shows an increase of the mass of the black hole while approaching the outer
apparent horizon. The increase is strongly dependent on numerical values of the model parameters and, due to the contribution of the scalar field, can lead to a significant increase in the gravitational mass of the central object. In the considered examples this increase can be of the order of 20\% as compared to the mass at infinity.

\subsubsection{Numerical fits of the solutions}

We have also obtained numerical fits for the three functions, $M_{eff}(\xi)$, $e^{\nu(\xi)}$ and $e^{-\lambda(\xi)}$, for all the combination of parameters $\phi_0=\{0.5, 0.6, 0.7, 0.8, 0.9\}$, $u_0=\{1,2,3,4,5\} \times 10^{-8}$, $\alpha=\{1.0, 1.5, 2.0, 2.5, 3.0\} \times 10^{-10}$, $\beta=\{ 1, 2, 3, 4, 5 \}\times 10^{-10}$.

For the mass function, we assume a general representation of the form
\begin{equation}\label{mass2}
M_{eff}(\xi)= A_{MH} + B_{MH} \xi + C_{MH} \xi^2 + D_{MH} \xi^3,
\end{equation}
where we further consider the coefficients $A_{MH}$, $B_{MH}$, $C_{MH}$ and $D_{MH}$ as functions $A_{MH}=A_{MH}\left(\phi_0, u_0, \alpha, \beta\right)$, $B_{MH}=B_{MH}\left(\phi_0, u_0, \alpha, \beta\right)$, $C_{MH}=C_{MH}\left(\phi_0, u_0, \alpha, \beta\right)$, and $D_{MH}=D_{MH}\left(\phi_0, u_0, \alpha, \beta \right)$, respectively. The explicit form of these coefficients is given below as
\begin{eqnarray}
A_{MH}&=&1 + 6.04244\times 10^7 \beta - 3.82478\times 10^6 u_0
	\nonumber\\
&&+\frac{1.024\times 10^6 u_0}{\phi_0}  +
 3.23921\times 10^6 \phi_0 u_0
 	\nonumber\\
 && - 1.09018\times 10^8 \alpha + 2.85266\times 10^{15}  \alpha \beta,
\end{eqnarray}
with an $R$ squared of $R^2=0.999961$,
\begin{eqnarray}
B_{MH}&=&1.787\times 10^7 \beta + 337899 u_0
\nonumber\\
&&
+ \frac{159812 u_0}{\phi_0}
	+159110 \phi_0 u_0  - 914980 \alpha
	\nonumber\\
&&- 4.13735\times 10^{16}  \alpha \beta,
\end{eqnarray}
with $R^2=0.997251$,
\begin{eqnarray}
C_{MH}&=&-4.65809\times 10^7 \beta - 468798 u_0 - \frac{310459. u_0}{\phi_0}
	\nonumber\\
&&- 477549 \phi_0 u_0 - 8.95984\times 10^6 \alpha
	\nonumber\\
&&+1.21263\times 10^{17}  \alpha \beta,
\end{eqnarray}
with $R^2=0.994383$, and finally
\begin{eqnarray}
D_{MH}&=&4.73217\times 10^7 \beta + 320946 u_0 + \frac{425716. u_0}{\phi_0}
	\nonumber\\
&&+701419. \phi_0 u_0 - 6.8347\times 10^6 \alpha
	\nonumber\\
&&- 1.04108\times 10^{17} \alpha \beta ,
\end{eqnarray}
with an $R$ squared of $R^2=0.99634$.

For the metric tensor coefficient  $e^{\nu(\xi)}$, we assume an analytical representation of the form
\begin{equation}
e^{\nu(\xi)}= A_{\nu H} + B_{\nu H} \xi + C_{\nu H} \xi^2 + D_{\nu H} \xi^3,
\end{equation}
where we further consider the coefficients $A$, $B$, $C$, $D$ as functions of the form
$A_{\nu H}=A_{\nu H}(\phi_0, u_0, \alpha, \beta)$, $B_{\nu H}=B_{\nu H}(\phi_0, u_0, \alpha, \beta)$, $C_{\nu H}=C_{\nu H}(\phi_0, u_0, \alpha, \beta)$, $D_{\nu H}=D_{\nu H}(\phi_0, u_0, \alpha, \beta)$, with the explicit forms of these functions  given below as
\begin{eqnarray}
A_{\nu H}&=&1 + 3.74259\times 10^7 \beta - 4377.59 u_0 + \frac{22095.4 u_0}{\phi_0}  	
	\nonumber\\
&&+57374.6 \phi_0 u_0 + 5.63505\times 10^7 \alpha
	\nonumber\\
&&-1.60231\times 10^{17} \alpha \beta ,
\end{eqnarray}
with an $R$ squared of $R^2=0.999996$,
\begin{eqnarray}
B_{\nu H}&=&-1 - 3.99701\times 10^7 \beta + 3.93268\times 10^6 u_0
	\nonumber\\
&&-\frac{858026 u_0}{\phi_0}  - 3.0797\times 10^6 \phi_0 u_0
 	\nonumber\\
&&+ 1.20215\times 10^8 \alpha  - 6.73634\times 10^{16} \alpha \beta ,
\end{eqnarray}
with $R^2=0.999962$,
\begin{eqnarray}
C_{\nu H}&=&-1.69096\times 10^7 \beta + 1.42883\times 10^6 u_0
	\nonumber\\
&&+ \frac{282037 u_0}{\phi_0}  + 14781.9 \phi_0 u_0 - 8.68885\times 10^7 \alpha
	\nonumber\\
&&+ 1.51073\times 10^{17}  \alpha \beta,
\end{eqnarray}
with $R^2=0.99119$, and finally
\begin{eqnarray}
D_{\nu H}&=&-2.6232\times 10^7 \beta - 1.209\times10^6 u_0
	\nonumber\\
&&-\frac{671312 u_0}{\phi_0}  - 656553 \phi_0 u_0 + 7.44748\times 10^7 \alpha
	\nonumber\\
&&-3.06922\times 10^{16}  \alpha \beta,
\end{eqnarray}
with an $R$ squared of $R^2=0.997681$.

For the metric tensor coefficient of $e^{-\lambda(\xi)}$, we adopt a functional representation  of the form
\begin{eqnarray}
e^{-\lambda(\xi)}&=&1-\xi M_{eff}(\xi)
	\nonumber\\
&=&1-(A_{MH} + B_{MH} \xi + C_{MH} \xi^2 + D_{MH} \xi^3)
	\nonumber\\
&=&1+A_{\lambda H}\xi+B_{\lambda H}\xi^2+C_{\lambda H}\xi^3+D_{\lambda H} \xi^4,
\end{eqnarray}
and we further consider the coefficients $A_{\lambda H}$, $B_{\lambda H}$, $C_{\lambda H}$, $D_{\lambda H}$ as functions of $\phi _0$ and $u_0$, respectively, so that  $A_{\lambda H}=A_{\lambda H}(\phi_0, u_0, \alpha, \beta)$, $B_{\lambda H}=B_{\lambda H}(\phi_0, u_0, \alpha, \beta)$, $C_{\lambda H}=C_{\lambda H}(\phi_0, u_0, \alpha, \beta)$, and $D_{\lambda H}=D_{\lambda H}(\phi_0, u_0, \alpha, \beta)$, respectively. The explicit forms of these functions are given below as
\begin{eqnarray}
A_{\lambda H}&=&-1 - 4.49859\times 10^7 \beta + 4.09247\times 10^6 u_0
	\nonumber\\
&&-\frac{966006 u_0}{\phi_0} - 3.19464\times 10^6 \phi_0 u_0
	\nonumber\\
&&+ 1.19154\times 10^8 \alpha - 5.17235\times 10^{16}  \alpha \beta ,
\end{eqnarray}
with an $R$ squared of $R^2=0.99996$,
\begin{eqnarray}
B_{\lambda H}&=&-9.47158\times 10^7 \beta - 1.36178\times 10^6 u_0
	\nonumber\\
&&-\frac{530466 u_0}{\phi_0}  - 626997 \phi_0 u_0
	\nonumber\\
&&-  3.64982\times 10^7 \alpha + 2.68594\times 10^{17} \alpha \beta,
\end{eqnarray}
with $R^2=0.993292$,
\begin{eqnarray}
C_{\lambda H}&=&1.78597\times 10^8 \beta + 1.6928\times 10^6 u_0
\nonumber\\
&&+\frac{1.09374\times 10^6 u_0}{\phi_0}  +
 1.71268\times 10^6 \phi_0 u_0
 	\nonumber\\
 &&+ 5.07146\times 10^7 \alpha - 4.84226\times 10^{17}  \alpha \beta ,
\end{eqnarray}
with $R^2=0.997681$, and finally
\begin{eqnarray}
D_{\lambda H}&=&-0.0292287 - 4.65552\times 10^7 \beta - 618525 u_0
	\nonumber\\
&&- \frac{932125 u_0}{\phi_0}  - 1.6095\times 10^6 \phi_0 u_0
	\nonumber\\
 &&+ 1.1466\times 10^8 \alpha - 3.58837\times 10^{16} \alpha \beta,
\end{eqnarray}
with an $R$ squared of $R^2=0.995662$.

\section{Thermodynamics of HMPG black holes}\label{sect5}

In the present analysis of the vacuum field equations in HMPG we have assumed that the mass function and lapse function $e^{\nu}$ depend only
on the radial coordinate. Hence  the spacetime is static and a timelike Killing vector $t^{\mu}$ exists \cite{Wald,Kunst}.  The definition of the surface gravity $\tilde{\kappa}$ for a static black hole that possesses a Killing horizon is given by \cite{Wald,Kunst}
\be
t^{\mu}\nabla _{\mu}t^{\nu}=t^{\nu}\tilde{\kappa}.
\ee
In the case of a static, spherically symmetric geometry that can be written as
\be
ds^2=-\tilde{\sigma} ^2 (r)f(r)c^2dt^2+\frac{dr^2}{f(r)}+r^2d\Omega ^2,
\ee
by adopting a suitable normalized Killing vector $t^{\mu}=\left(1/\tilde{\sigma}_{\infty},0,0,0\right)$, the surface gravity of the black hole can be obtained as \cite{Kunst}
\be
\tilde{\kappa}=\left(\frac{\tilde{\sigma} _{hor}}{\tilde{\sigma} _{\infty}}\right)\frac{c^4}{4GM_{hor}}\left.\left[1-\frac{2GM'(r)}{c^2}\right]\right|_{hor},
\ee
where the subscript {\it hor} indicates that the evaluation of all physical quantities must be performed  on the outer
apparent horizon. For $\sigma \equiv 1$, and $M={\rm constant}$, we reobtain the well-known result of the surface gravity of a Schwarzschild black hole, $\tilde{\kappa}=c^4/4GM_{hor}$ \cite{Wald}. The temperature $T_{BH}$ of the black hole is defined as
\be
T_{BH}=\frac{\hbar}{2\pi ck_B} \tilde{\kappa},
\ee
where $k_B$ is Boltzmann's constant. In the dimensionless variables introduced in Eq.~(\ref{dimvar}) we obtain the temperature of the black hole as
\be
T_{BH}=T_H\frac{1}{M_{eff}\left(\xi _S\right)}\left.\left(1+\xi ^2\frac{dM_{eff}\left(\xi\right)}{d\xi}\right)\right|_{\xi =\xi _S},
\ee
where
\be
T_H=\frac{\hbar c^3}{8\pi Gk_BnM_{\odot}}.
\ee

By taking into account the representation of the effective mass as given by Eqs.~(\ref{mass1}) and (\ref{mass2}), we obtain for the temperature of a HMPG black hole, the expression
\bea
T_{BH}\left(\xi _S\right)&=&T_H\left.\frac{1+\xi ^2\left(B+2C\xi+3D\xi^2\right)}{A+B\xi +C\xi^2+D\xi ^3}\right|_{\xi =\xi _S}
	\nonumber\\
&=&T_H\left.\theta \left(\xi\right)\right|_{\xi =\xi _S}.
\eea

\begin{figure}[!htb]
\centering
\includegraphics[scale=0.65]{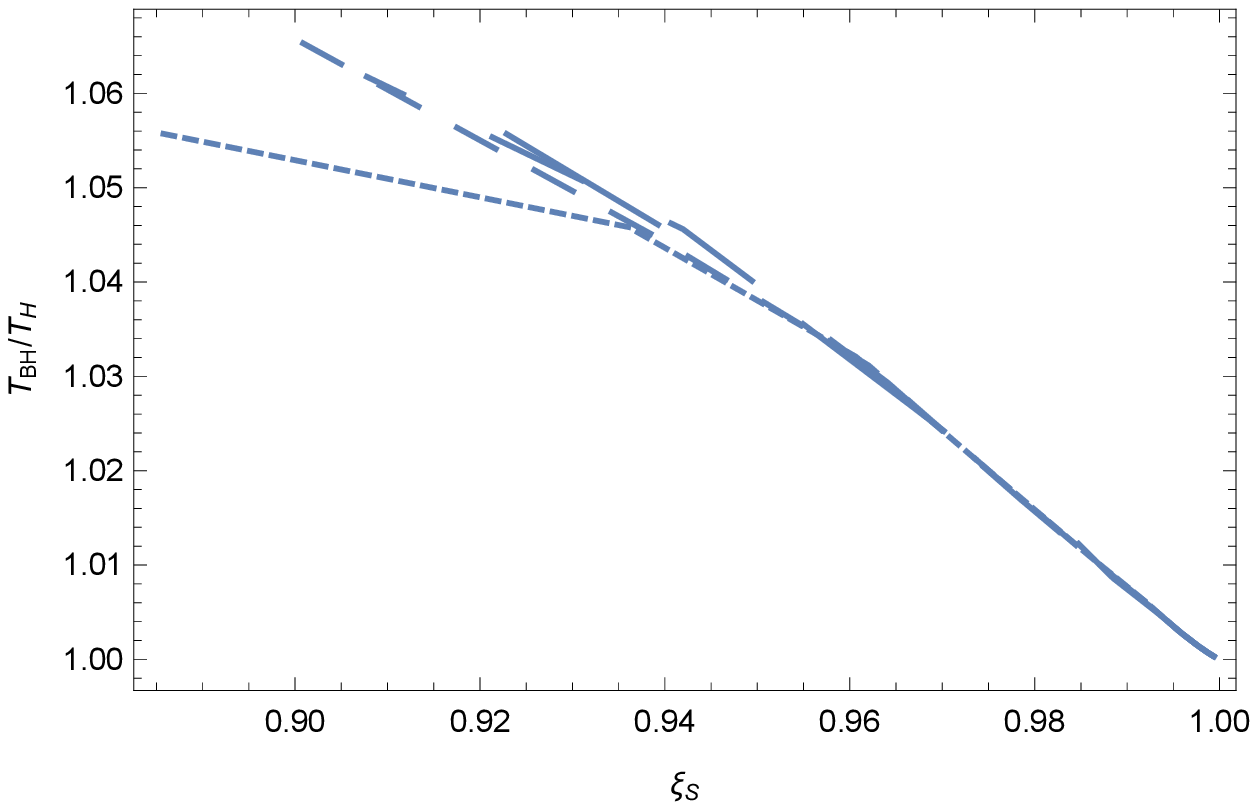}
\caption{Variation of the black hole temperature $T_{BH}\left(\xi _S\right)/T_H$
as a function of the horizon radius $\xi _S$ of a HMPG black hole in the absence of a scalar field potential, $V=0$,  for $u_0\in \left[2.2\times 10^{-10},1.1264\times 10^{-7}\right]$, and for different values of $\phi _0$: $\phi _0=0.15$ (solid curve), $\phi _0=0.225$ (dotted curve), $\phi _0=0.3375$ (short dashed curve), $\phi _0=0.50625$ (dashed curve), respectively.}
\label{fig7}
\end{figure}

For the zero potential case $V=0$ the variation of the horizon temperature of HMPG black holes is represented in Fig.~\ref{fig7}. Explicit numerical values of the $T_{BH}\left(\xi _S\right)/T_H$ ratio are presented in Table~\ref{table5}.

\begin{table}
\begin{center}
\begin{tabular}{ |c|c|c|c| }
 \hline
 $\phi_0$/$u_0$ & $0.15$ & $0.225$ & $0.3375$ \\
 \hline
 $2.2 \times 10^{-10}$ & 1.0004 & 1.00038 & 1.00035 \\
 $1.76 \times 10^{-9}$ & 1.00316 & 1.00298 & 1.00273 \\
 $1.408 \times 10^{-8}$ & 1.01904 & 1.01825 & 1.017 \\
 $1.1264 \times 10^{-7}$ & 1.05587 & 1.0653 & 1.06176 \\
 \hline
\end{tabular}
\caption{Selected numerical values of the black hole temperature $T_{BH}\left(\xi _S\right)/T_H$ in the absence of a scalar field potential, $V=0$, for $\phi_0=\{ 0.15, 0.225, 0.3375 \}$ and $u_0=\{ 2.2 \times 10^{-10}, 1.76 \times 10^{-9}, 1.408 \times 10^{-8}, 1.1264 \times 10^{-8} \}$.}\label{table5}
\end{center}
\end{table}

The specific heat $C_{BH}$ of the black hole can be obtained as
\bea
C_{BH}&=&\frac{dM}{dT_{BH}}=\left.\frac{dM}{dr}\frac{dr}{dT_{BH}}\right|_{r=r_{hor}}
	\nonumber\\
&=&\frac{nM_{\odot}}{T_H} \left.\frac{dM_{eff}\left(\xi \right)}{d\xi}\frac{d\xi}{d\theta }\right|_{\xi =\xi _S}.
\eea

\begin{figure}[!htb]
\centering
\includegraphics[scale=0.65]{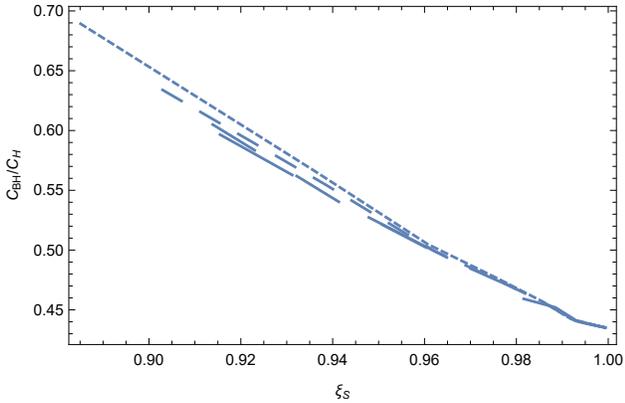}
\caption{Variation of the specific heat $C_{BH}\left(\xi _S\right)/C_H$
as a function of the horizon radius $\xi _S$ of a HMPG black hole in the absence of a scalar field potential, $V=0$,  for $u_0\in \left[2.2\times 10^{-10},1.1264\times 10^{-7}\right]$, and for different values of $\phi _0$: $\phi _0=0.15$ (solid curve), $\phi _0=0.225$ (dotted curve), $\phi _0=0.3375$ (short dashed curve), $\phi _0=0.50625$ (dashed curve), respectively.}
\label{fig8}
\end{figure}

Hence for the specific heat of a black hole in HMPG we obtain the general expression
\begin{widetext}
\be
C_{BH}\left(\xi _S\right)=C_H\frac{\left[B+\xi  (2 C+3 D \xi )\right] \left\{A+\xi  \left[B+\xi  \left(C+D \xi \right)\right]\right\}^2}{
   B \left[2 A \xi +4 \xi ^3 \left(C+2 D \xi \right)-1\right]+\xi  \left[C \left(6 A
   \xi +6 D \xi ^4-2\right)+3D \xi  \left(4 A \xi +D \xi ^4-1\right)+2
   C^2 \xi ^3\right]+B^2 \xi ^2},
\ee
\end{widetext}
where we have denoted $C_H=nM_{\odot}/T_H$. The variation of the specific heat of the HMPG black holes as a function of the dimensionless horizon radius is represented, for the zero potential case $V=0$, in Fig.~\ref{fig8}. Exact numerical values of the ratio $C_{BH}\left(\xi _S\right)/C_H$ for different values of $\phi _0$ and $u_0$ are presented in Table~\ref{table6}.

\begin{table}
\begin{center}
\begin{tabular}{ |c|c|c|c| }
 \hline
$\phi _0$/$u_0$ & $0.15$ & $0.225$ & $0.3375$ \\
 \hline
 $2.2 \times 10^{-10}$ & 0.435307 & 0.435248 & 0.43189 \\
 $1.76 \times 10^{-9}$ & 0.43887 & 0.438458 & 0.438062 \\
 $1.408 \times 10^{-8}$ & 0.475718 & 0.472819 & 0.470153 \\
 $1.1264 \times 10^{-7}$ & 0.68967 & 0.638562 & 0.619363 \\
 \hline
\end{tabular}
\caption{Numerical values of the specific heat $C_{BH}\left(\xi _S\right)/C_H$
of a HMPG black hole in the absence of a scalar field potential, $V=0$,  for $\phi_0=\{ 0.15, 0.225, 0.3375 \}$ and $u_0=\{ 2.2 \times 10^{-10}, 1.76 \times 10^{-9}, 1.408 \times 10^{-8}, 1.1264 \times 10^{-8} \}$.}\label{table6}
\end{center}
\end{table}

The entropy $S_{BH}$ of the black hole is given by
\bea
S_{BH}&=&\int_{r_{in}}^{r_{hor}}{\frac{dM}{T_{BH}}}=\int_{r_{in}}^{r_{hor}}{\frac{1}{T_{BH}}\frac{dM}{dr}dr},
\eea
or, equivalently,
\bea
S_{BH}\left(\xi _S\right)=C_H\int_0^{\xi _S}{\frac{1}{\theta \left(\xi \right)}\frac{dM_{eff}\left(\xi\right)}{d\xi}d\xi}.
\eea

\begin{figure}[!htb]
\centering
\includegraphics[scale=0.65]{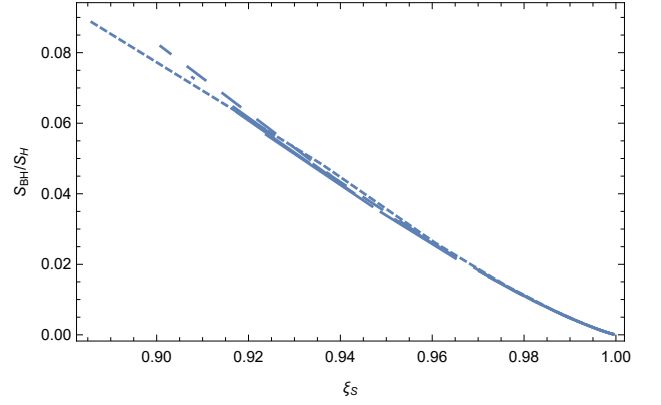}
\caption{Variation of  the entropy $S_{BH}\left(\xi _S\right)/C_H$  of a HMPG black hole as a function of the horizon radius $\xi _S$ in the absence of a scalar field potential, $V=0$,  for $u_0\in \left[2.2\times 10^{-10},1.1264\times 10^{-7}\right]$, and for different values of $\phi _0$: $\phi _0=0.15$ (solid curve), $\phi _0=0.225$ (dotted curve), $\phi _0=0.3375$ (short dashed curve), $\phi _0=0.50625$ (dashed curve), respectively.}
\label{fig9}
\end{figure}

 The variation  as a function of the dimensionless horizon radius  $\xi _S$ of the entropy of the HMPG black holes is represented, for the zero potential case $V=0$, in Fig.~\ref{fig9}. Selected values of the ratio $S_{BH}\left(\xi _S\right)/C_H$ for different values of $\phi _0$ and $u_0$ are presented in Fig.~\ref{table7}.

\begin{table}
\begin{center}
\begin{tabular}{ |c|c|c|c| }
 \hline
$\phi _0$/$u_0$  & $0.15$ & $0.225$ & $0.3375$ \\
 \hline
 $2.2 \times 10^{-10}$ & $2.414\times 10^{-4}$ & $2.267\times 10^{-4}$ & $2.076\times 10^{-4}$ \\
 $1.76 \times 10^{-9}$ & $1.905\times 10^{-3}$ & $1.790\times 10^{-3}$ & $1.641\times 10^{-3}$ \\
 $1.408 \times 10^{-8}$ & $1.384\times 10^{-2}$ & $1.303\times 10^{-2}$ & $1.199\times 10^{-2}$ \\
 $1.126 \times 10^{-7}$ & $8.938\times 10^{-2}$ & $8.181\times 10^{-2}$ & $7.299\times 10^{-2}$ \\
 \hline
\end{tabular}
\caption{Numerical values of the entropy $S_{BH}\left(\xi _S\right)/C_H$  of a HMPG black hole in the absence of a scalar field potential, $V=0$,  for $\phi_0=\{ 0.15, 0.225, 0.3375 \}$ and $u_0=\{ 2.2 \times 10^{-10}, 1.76 \times 10^{-9}, 1.408 \times 10^{-8}, 1.1264 \times 10^{-8} \}$.}\label{table7}
\end{center}
\end{table}

The black hole luminosity due to the Hawking evaporation can be obtained as
\be
L_{BH}=-\frac{dM}{dt}=-\sigma A_{BH}T_{BH}^4,
\ee
where $\sigma $ is a model dependent parameter, and $A_{BH}=4\pi r_{hor}^2$ is the area of the event horizon. Hence for the black hole evaporation time $\tau $ we find
\bea
\hspace{-0.8cm}\tau &=&\int_{t_{in}}^{t_{fin}}{dt}=-\frac{1}{4\pi \sigma}\int_{t_{in}}^{t_{fin}}{\frac{dM}{r_{hor}^2T_{BH}^4}},
\eea
or equivalently,
\bea
\tau \left(\xi _S\right)=-\tau _H\int_0^{\xi _S}{\frac{1}{\xi ^2\theta ^4\left(\xi\right)}\frac{dM_{eff}\left(\xi\right)}{d\xi }d\xi},
\eea
where we have denoted
\be
\tau _H=\frac{c^4}{8\pi G^2\sigma nM_{\odot}T_{BH}^4}.
\ee

\begin{figure}[!htb]
\centering
\includegraphics[scale=0.65]{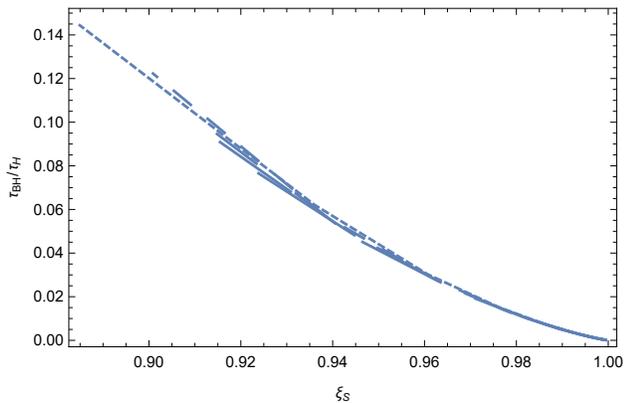}
\caption{Variation of the evaporation time $\tau_{BH}\left(\xi _S\right)/\tau _H$ as a function of the horizon radius $\xi _S$  of a HMPG black hole in the absence of a scalar field potential, $V=0$,  for $u_0\in \left[2.2\times 10^{-10},1.1264\times 10^{-7}\right]$, and for different values of $\phi _0$: $\phi _0=0.15$ (solid curve), $\phi _0=0.225$ (dotted curve), $\phi _0=0.3375$ (short dashed curve), $\phi _0=0.50625$ (dashed curve), respectively.}
\label{fig10}
\end{figure}

 The variation  of the Hawking evaporation time as a function of the dimensionless horizon radius  $\xi _S$ of the HMPG black holes is represented, for the zero potential case $V=0$, in Fig.~\ref{fig10}. Explicit exact numerical values of the evaporation time $\tau_{BH}\left(\xi _S\right)/\tau _H$ are presented, for different values of $\phi _0$ and $u_0$,  in Table~\ref{table8}.

\begin{table}
\begin{center}
\begin{tabular}{ |c|c|c|c| }
 \hline
$\phi _0$/$u_0$ & $0.15$ & $0.225$ & $0.3375$ \\
 \hline
 $2.2 \times 10^{-10}$ & $2.421\times 10^{-4}$ & $2.273\times 10^{-4}$ & $2.082\times 10^{-4}$ \\
 $1.76 \times 10^{-9}$ & $1.938\times 10^{-3}$ & $1.819\times 10^{-3}$ & $1.666\times 10^{-3}$ \\
 $1.408 \times 10^{-8}$ & $1.513\times 10^{-2}$ & $1.419\times 10^{-2}$ & $1.299\times 10^{-2}$ \\
 $1.126 \times 10^{-7}$ & $1.444\times 10^{-1}$ & $1.222\times 10^{-1}$ & $1.058\times 10^{-1}$ \\
 \hline
\end{tabular}
\caption{Numerical values  of the evaporation time $\tau_{BH}\left(\xi _S\right)/\tau _H$  of a HMPG black hole in the absence of a scalar field potential, $V=0$,  for $\phi_0=\{ 0.15, 0.225, 0.3375 \}$ and $u_0=\{ 2.2 \times 10^{-10}, 1.76 \times 10^{-9}, 1.408 \times 10^{-8}, 1.1264 \times 10^{-8} \}$.}\label{table8}
\end{center}
\end{table}

\section{Discussions and final remarks}\label{sect6}

In the present paper, we have investigated the possible existence of black hole type structures in the framework of the HMPG theory, by considering the simplest case, corresponding to a vacuum static and spherically symmetric geometry. Even within this simple theoretical model the field equations of the theory become extremely complicated, and therefore in order to obtain solutions of the field equations one must resort to numerical methods. To this effect, we have reformulated the static spherically symmetric Einstein field equations in their scalar-tensor representation in a dimensionless form, and introduced the inverse of the radial coordinate as the independent variable. This representation allows an easier numerical integration procedure, which also requires fixing the numerical values of the scalar field, of its derivative, and of the effective mass at infinity. The appearance of a singular behavior in the field equations or, more exactly, in the behavior of the metric tensor coefficients, is interpreted as indicating the presence of an event horizon and, consequently, of a black hole type object. The mass of the black hole is given by the effective mass of the model, which represents the total contribution of the ordinary mass of the black hole plus the contribution from the scalar field.

We have considered the solutions of the gravitational field equations of the HMPG theory for two choices of the scalar field potential $V\left(\phi\right)$, corresponding to the cases of the vanishing potential, and of the Higgs-type potential, respectively. In both cases our results indicate the formation of an event horizon, and consequently of black holes. The position of the event horizon depends on the values of the scalar field and of its derivative at infinity (the initial conditions), indicating the existence of a complex relation between scalar field and black hole properties. For the zero potential case and for particular scalar field initial conditions the event  horizon can be located at distances of the order of 0.7 of the standard Schwarzschild radius,
indicating the formation of more compact black holes as compared to standard general relativity. In the case of the Higgs-type potential the position of the event horizon is also strongly dependent on the parameters $\alpha $ and $\beta$ of the potential, indicating a multi-parametric dependence of the black hole properties. In all the cases studied it turns out that the numerical results can be fitted well by some simple analytical functions. In the zero potential case the metric function $e^{\nu(r)}$ can be described by a function of the type $e^{\nu}=1-B/r-C/r^2+D$, with $A,B,C,D$ constants that depend on the initial conditions at infinity. The metric tensor component $e^{-\lambda}$ also contains a term proportional to $1/r^3$. Similar simple analytic representations can describe the numerical results for the case of the Higgs-type potential. These analytical representations are extremely useful in the study of the thermodynamic properties of the HMPG black holes, as well as the dynamics and motion of matter particles around them. In particular, they may be used for the study of the electromagnetic properties of accretion disks that form around black holes, and which could allow discriminating this type of theoretical objects from their general relativistic counterparts, and for obtaining some constraints on the model parameters.

We have also investigated in detail the thermodynamic properties of the obtained numerical black hole solutions. One of the essential and interesting physical properties of black holes is their Hawking temperature. As compared to the standard general relativistic Hawking temperature, the horizon temperature of the HMPG black holes shows a strong dependence on the initial conditions at infinity, and the properties of the scalar field potential. As one can see from Fig.~\ref{fig7}, a decrease in the horizon radius leads to a higher black hole temperature, which, in the case of the specific initial conditions considered in Fig.~\ref{fig7}, is of the order of 10\%, as compared to the standard general relativistic case. Similar effects appear for the specific heat, entropy and evaporation time of the HMPG black holes, all these quantities being strongly dependent on the initial conditions of the scalar field at infinity. In particular, the black hole  evaporation times may be very different in HMPG as compared to standard general relativity. Of course our results on the thermodynamics of black holes, obtained for the zero potential case and for a limited set of initial conditions at infinity may be considered on qualitative nature only. But even at this level they indicate the complexity of the behavior of the HMPG black holes, and of the interesting physics related to them.

Black hole solutions are also well known in standard scalar field models. For a nonminimally coupled scalar field such exact analytical solutions have been obtained and studied a long time ago in \cite{Fisher, Bergmann, Newman, Bron, Sol, Tur} (for a recent review nonsingular static, spherically symmetric solutions of general relativity with minimally coupled scalar fields see \cite{Bronrev}). These solutions have been generally obtained in the Einstein frame, in which there is no coupling between the scalar field and the Ricci scalar. On the other hand because of the specific coupling between the scalar field and the Ricci scalar, the HMPG theory appears to be naturally formulated in the Jordan frame. Despite its superficial resemblance with the Brans-Dicke theory with coupling $w=-3/2$, there are fundamental differences between the HMPG theory and scalar field models in the Einstein or Jordan frames. One such important difference appears in the zero potential case. While in the standard scalar field models the solutions with zero potential have in general no horizons, our investigations show that this is generally not the case in the HMPG theory, where even in the zero potential case the formation of ordinary black holes occur. In the standard scalar field models such a situation may occur for solutions admitting a conformal continuation, meaning that a singularity in the Einstein-frame manifold maps to a regular surface in the Jordan frame, and the solution is then continued beyond this surface \cite{Bronn1}.

All possible types of spacetime causal structures that can appear in static, spherically symmetric configurations of a self-gravitating minimally coupled scalar field $\phi$ in general relativity, with an arbitrary potential $V(\phi)$, were considered in \cite{Bronn2}. It was first shown that a variable scalar field does not modify the possible structures with a constant scalar field. Moreover,   in general relativistic scalar field models with arbitrary  $V(\phi)$ there are no regular black holes with flat or AdS asymptotics. It also follows that the possible globally regular, asymptotically flat solutions are solitons with a regular center, without horizons and with at least partly negative potentials $V(\phi)$. For a similar discussion of higher dimensional models see \cite{Bronn3}.  These results cannot be recovered in HMPG theory, in which in the case of Higgs-like potentials black hole solutions presenting an event horizon exist. In fact our numerical investigations did not reveal the presence of any globally regular solution.

An important result in black hole physics is the no-hair theorem \cite{Bek, Bek1, Adler, Bek2}, stating that asymptotically flat black holes cannot possess external nontrivial  scalar fields with non-negative field potential $V (\phi)$. The results obtained in the present paper indicate that the no-hair theorem in its standard formulation cannot be extended to HMPG theory. All the considered black hole solutions are asymptotically flat, and scalar fields with positive potentials exist around them. However, the question if such structures result from a particular choice of the scalar field potentials and of the model parameters, or they are intrinsic properties of the theory deserves further investigation.

HMPG black holes may present a much richer theoretical structure, properties and variability, associated with an equally rich external dynamics, as compared with the standard general relativistic black holes. These properties are related to the presence of the intricate coupling between the scalar field, geometry and matter, which leads to very complex, strongly nonlinear, field equations. These new effects can also lead to some specific astrophysical signatures and imprints, whose observational detection could lead to new perspectives in gravitational physics and astrophysics. The possible astrophysical/observational implications  of the existence of HMPG black holes  will be considered
in a future publication.

\section*{Acknowledgments}

B.D. acknowledges financial support from PNIII STAR ACRONIM ASTRES: Centre
of Competence For Planetary Sciences, Nr. 118/14.11.2016. F.S.N.L.
acknowledges financial support of the Funda\c{c}\~{a}o para a Ci\^{e}ncia e
Tecnologia through the research grants No. UID/FIS/04434/2013, No.~PEst-OE/FIS/UI2751/2014 and No. PTDC/FIS-OUT/29048/2017. T.H. would like to
thank the Yat-Sen School of the Sun Yat-Sen University in Guangzhou, P. R.
China, for the kind hospitality offered during the preparation of this work.


\appendix
\section{The dimensionless representation of the geometric and physical quantities}\label{Appa}

In the following we present the explicit relations for the transformation of the dimensional quantities to dimensionless ones under the scaling introduced in Eqs.~(\ref{dimvar}). They are as follows:

\begin{equation}
\frac{dm_{eff}}{dr}=nM_{\odot}\frac{dM_{eff}}{d\eta }\frac{c^{2}}{%
2GM_{\odot}n}=\frac{c^{2}}{2G}\frac{dM_{eff}}{d\eta },
\end{equation}
\begin{equation}
\frac{2Gm_{eff}(r)}{c^{2}r}=\frac{2GnM_{\odot}M_{eff}\left( \eta \right) }{%
c^{2}\frac{2GM_{\odot}}{c^{2}}n\eta }=\frac{M_{eff}\left( \eta \right) }{%
\eta },
\end{equation}
\begin{equation}
\frac{du}{dr}=\frac{c^{2}}{2GM_{\odot}n}\frac{dU}{d\eta }\frac{c^{2}}{%
2GM_{\odot}n}=\left( \frac{c^{2}}{2GM_{\odot}n}\right) ^{2}\frac{dU}{d\eta },
\end{equation}
\begin{equation}
ur=\frac{c^{2}}{2GM_{\odot}n}U\left( \eta \right) \frac{2GM_{\odot}}{c^{2}}%
n\eta =\eta U\left( \eta \right) ,
\end{equation}
\bea
m_{eff}u&=&nM_{\odot}M_{eff}\left( \eta \right) \frac{c^{2}}{2GM_{\odot}n}%
U\left( \eta \right)
	\nonumber\\
&=&\frac{c^{2}}{2G}M_{eff}\left( \eta \right) U\left(
\eta \right) ,
\eea
\begin{equation}
\frac{u}{r}=\frac{c^{2}}{2GM_{\odot}n}U\left( \eta \right) \frac{c^{2}}{%
2GM_{\odot}n\eta }=\left( \frac{c^{2}}{2GM_{\odot}n}\right) ^{2}\frac{%
U\left( \eta \right) }{\eta }.
\end{equation}

\end{document}